\documentclass[onecollarge,runningheads]{svjour2}
\smartqed  
\usepackage{graphicx}

 \usepackage{mathptmx}      
\usepackage{graphicx}
 \usepackage{mathptmx}      
    \usepackage{bm}

    \usepackage{dcolumn}

    \usepackage{amsmath}

    \usepackage{amssymb}
    \usepackage{revsymb}

\newcommand\beq{\begin{equation}}
\newcommand\eeq{\end{equation}}
\newcommand\beqa{\begin{eqnarray}}
\newcommand\eeqa{\end{eqnarray}}
\newcommand{\nn}{\nonumber\\}

\newcommand{\rr}{k}

\newcommand{\kk}{\alpha}

\newcommand{\alphak}{\vartheta}

\newcommand{\gw}{w}

\newcommand{\ww}{\zeta}

\newcommand{\omegak}{\omega}

\newcommand{\gammak}{\varpi}

\journalname{Continuum Mechanics and Thermodynamics}
\begin{document}

\title{Solutions of the moment hierarchy in the kinetic theory of Maxwell models}



\author{Andr\'es Santos
}


\institute{A. Santos \at
              Departamento de F\'{\i}sica, Universidad
de Extremadura, E-06071 Badajoz, Spain \\
              Tel.: +34-924289540\\
              Fax: +34-924289651\\
              \email{andres@unex.es}
}


\maketitle

\begin{abstract}
In the Maxwell interaction model the collision rate is independent of the relative velocity of the colliding pair and, as a consequence, the collisional moments are bilinear combinations of velocity moments of the same or lower order. In general, however, the drift term of the Boltzmann equation couples moments of a given order to moments of a higher order, thus preventing the solvability of the moment hierarchy, unless  approximate closures are introduced. On the other hand, there exist a number of states where the moment hierarchy can be recursively solved, the solution generally exposing non-Newtonian properties. The aim of this paper is to present an overview of results pertaining to some of those states, namely  the planar Fourier flow (without and with a constant gravity field), the planar Couette flow,  the force-driven Poiseuille flow, and the uniform shear flow.
\end{abstract}

\section{Introduction}
\label{sec1}
As is well known, the classical kinetic theory of low-density gases began to be established as a mathematically sound statistical-physical theory with the work of James Clerk Maxwell  (1831--1879) \cite{B65,B76,B03}. Apart from obtaining the velocity distribution function at equilibrium (first in 1860 and then, in a more rigorous way, in 1867), Maxwell derived in 1866 and 1867 the transfer  equations characterizing the rate of change of any quantity (such as mass, momentum, or energy) which can be defined in terms of molecular properties. This paved the way to Ludwig Boltzmann (1844--1906) in the derivation of his celebrated equation (1872) for the rate of change of the velocity distribution itself. As a matter of fact, the Boltzmann equation is formally equivalent to Maxwell's infinite set  of transfer equations.

In his work entitled ``On the Dynamical Theory of Gases,'' Maxwell departs from the hard-sphere model and writes \cite{B65,B03}
\begin{quotation}
``In the present paper I propose to consider the molecules of a gas, not as elastic spheres of definite radius, but as small bodies or groups of smaller molecules repelling one another with a force whose direction always passes very nearly through the centres of gravity of the molecules, and whose magnitude is represented very nearly by some function of the distance of the centres of gravity. I have made this modification of the theory in consequence of the results of my experiments on the viscosity of air at different temperatures, and I have deduced from these experiments that the repulsion is inversely as the \emph{fifth} power of the distance.''
\end{quotation}
Maxwell realized that the hypothesis of a force inversely proportional to the fifth power of the distance, or, equivalently, of an interaction potential $\varphi(r)\propto r^{-4}$, makes the nonequilibrium distribution function $f$ enter the transfer equations in such a way that the transport coefficients can be evaluated without the need of knowing the detailed form of $f$. His results with this interaction model showed that the shear viscosity was proportional to the absolute temperature $T$, whereas in the case of hard spheres it is proportional to the square root of the temperature. As said in the preceding quotation, Maxwell himself carried out a series of experiments to measure the viscosity of air as a function of temperature. He writes  \cite{B65,B03}
\begin{quotation}
``I have found by experiment that the coefficient of viscosity in a given gas is independent of the density, and proportional to the absolute temperature, so that if $ET$ be the viscosity, $ET\propto  p/\rho$.''
\end{quotation}
Modern measurements of the viscosity of air show that it is approximately proportional to $T^{0.76}$ over a wide range of temperatures \cite{A53,A89}, so that the actual power is intermediate  between the values predicted by the hard-sphere and Maxwell models.

In any case, the adoption of Maxwell's interaction model $\varphi(r)\propto r^{-4}$ allows one to extract some useful information from the nonlinear Boltzmann equation, or its associated set of moment equations, for states far from equilibrium. Apart from its intrinsic interest, exact results derived for Maxwell molecules are important as benchmarks to assess approximate moment methods, model kinetic equations, or numerical algorithms (deterministic or stochastic) to solve the Boltzmann equation. Moreover, it turns out that many consequences of the Boltzmann equation for Maxwell molecules, when properly rescaled with respect to the collision frequency, can be (approximately) extrapolated to more general interaction potentials \cite{GS03}.

The aim of this paper is to review a few examples of nonequilibrium  states whose hierarchy of moment equations can be recursively solved for the Maxwell model. The structure and main properties of the moment equations for Maxwell molecules are recalled in Section \ref{sec2}. This is followed by a description of the solution for the planar Fourier flow (Section \ref{sec3}),  the planar Couette flow (Section \ref{sec4}), the force-driven Poiseuille flow (Section \ref{sec5}), and the uniform shear flow (Section \ref{sec6}). The paper is closed with some concluding remarks in Section \ref{sec7}.

\section{Moment equations for Maxwell molecules}
\label{sec2}

\subsection{The Boltzmann equation}
Let us consider a dilute gas made of particles of mass $m$ interacting via a short-range pair potential $\varphi(r)$. The relevant statistical-mechanical description of the gas is conveyed by the one-particle velocity distribution function $f(\mathbf{r},\mathbf{v},t)$. The  number density  $n(\mathbf{r},t)$, the flow velocity  $\mathbf{u}(\mathbf{r},t)$, and the temperature $T(\mathbf{r},t)$ are related to $f$ through
\begin{equation}
n(\mathbf{r},t)=\int d\mathbf{v} \,f(\mathbf{r},\mathbf{v},t),
\label{1.2:1}
\end{equation}
\begin{equation}
\mathbf{u}(\mathbf{r},t)=\frac{1}{n(\mathbf{r},t)}\int d\mathbf{v}\, \mathbf{v} f(\mathbf{r},\mathbf{v},t),
\label{1.2:2}
\end{equation}
\begin{equation}
n(\mathbf{r},t)k_BT(\mathbf{r},t)=p(\mathbf{r},t)=\frac{m}{3}\int d\mathbf{v}\, {V}^2(\mathbf{v},\mathbf{r},t) f(\mathbf{r},\mathbf{v},t),
\label{1.2:3}
\end{equation}
where  $k_B$ is the  Boltzmann constant and $\mathbf{V}(\mathbf{v},\mathbf{r},t)=\mathbf{v}- \mathbf{u}(\mathbf{r},t)$ is the so-called peculiar velocity.
The fluxes of momentum and energy are measured by the pressure (or stress) tensor $\mathsf{P}$ and the heat flux vector $\mathbf{q}$, respectively. Their expressions are
\begin{equation}
\mathsf{P}(\mathbf{r},t)=m\int d\mathbf v\, \mathbf{V}(\mathbf{v},\mathbf{r},t)\mathbf{V}(\mathbf{v},\mathbf{r},t) f(\mathbf{r},\mathbf{v},t),
\label{1.3:12}
\end{equation}
\begin{equation}
\mathbf{q}(\mathbf{r},t)=\frac{m}{2}\int d\mathbf v\, V^2(\mathbf{v},\mathbf{r},t)\mathbf{V}(\mathbf{v},\mathbf{r},t) f(\mathbf{r},\mathbf{v},t).
\label{1.3:15}
\end{equation}

The time  evolution of the velocity distribution function is governed by the Boltzmann equation \cite{CC70,DvB77}
\beq
\frac{\partial }{\partial t}f=-\mathbf{v}\cdot \nabla f-
\frac{\partial}{\partial \mathbf{v}}\cdot \left(\frac{\mathbf{F}}{m}f\right)+J[\mathbf{v}|f,f].
\label{2.1}
\eeq
The first and second terms on the right-hand side represent the rate of change of $f$ due to the free motion and to the possible action of an external force $\mathbf{F}$, respectively. In general, such a force can be non-conservative and velocity-dependent, and in that case it must appear to the right of the operator ${\partial}/{\partial \mathbf{v}}$. The last term on the right-hand side is the fundamental one. It gives the rate of change of $f$ due to the interactions among the particles treated as successions of local and instantaneous binary collisions. The explicit form of the (bilinear) collision operator $J[\mathbf{v}|f,f]$ is
\beq
J[\mathbf{v}|f,f]=\int d\mathbf{v}_1\int d\Omega\,
\gw B(\gw,\chi) \left[f(\mathbf{v}')f(\mathbf{v}_1')-f(\mathbf{v})f(\mathbf{v}_1)\right],
\label{2.2}
\eeq
where $\mathbf{\gw}\equiv\mathbf{v}-\mathbf{v}_1$ is the pre-collisional relative velocity,
\beq
\mathbf{v}'=\mathbf{v}-(\mathbf{\gw}\cdot \widehat{\bm{\sigma}})\widehat{\bm{\sigma}},
\quad
\mathbf{v}_1'=\mathbf{v}_1+(\mathbf{\gw}\cdot \widehat{\bm{\sigma}})\widehat{\bm{\sigma}}
\label{2.2b}
\eeq
are the post-collisional velocities, $\widehat{\bm{\sigma}}$ is a unit vector pointing on the apse line (i.e., the line joining the centers of the two particles at their closest approach), $\chi$ is the scattering angle (i.e., the angle between the post-collisional relative velocity $\mathbf{\gw}'\equiv\mathbf{v}'-\mathbf{v}_1'$ and $\mathbf{\gw}$), and $d\Omega=4|\widehat{\mathbf{\gw}}\cdot \widehat{\bm{\sigma}}|d\widehat{\bm{\sigma}}$ is an element of solid angle about the direction of $\mathbf{\gw}'$. Finally, $B(\gw,\chi)$ is the differential cross section \cite{G59}. This is the only quantity that
depends on the choice of the potential $\varphi(r)$:
\beq
B(\gw,\chi)=\frac{b(\gw,\chi)}{\sin\chi}\left| \frac{\partial b(\gw,\chi)}{\partial\chi}\right|,
\label{2.3}
\eeq
where the impact parameter $b(\gw,\chi)$ is obtained from inversion of
\beq
\chi(\gw,b)=\pi -2\int_{r_0(\gw,b)}^\infty dr\frac{b/r^2}{\left[1-(b/r)^2-4\varphi(r)/m\gw^2\right]^{1/2}},
\label{2.4}
\eeq
$r_0(\gw,b)$ being the distance at closest approach, which is given as the root of $1-(b/r)^2-4\varphi(r)/m\gw^2$.

In the special case of  inverse power-law (IPL) repulsive potentials of the form  $\varphi(r)=\varphi_0 (\sigma/r)^{\ww}$,
the  scattering
angle $\chi$ depends on both $b$  and $\gw$ through the scaled dimensionless parameter
$\beta=(b/\sigma)(m\gw^2/2\ww \varphi_0)^{1/\ww}$ ($\varphi_0$, $\sigma$, and $\ww$ being constants), namely
\begin{equation}
\chi(\beta)=\pi-2\int_0^{\beta_0(\beta)} d\beta' \,\left[1-{\beta'}^2-\frac{2}{\ww}
\left(\frac{\beta'}{\beta}\right)^\ww\right]^{-1/2},
\label{2.5}
\end{equation}
where $\beta_0(\beta)$ is the root of the quantity enclosed by brackets. For this class of potentials the differential cross section has the scaling form
\beq
B(\gw,\chi)=\sigma^2 \left(\frac{2\ww \varphi_0}{m\gw^2}\right)^{2/\ww}\mathcal{B}(\chi),\quad \mathcal{B}(\chi)=\frac{\beta(\chi)}{\sin\chi}\left| \frac{d \beta(\chi)}{d\chi}\right|.
\label{2.6}
\eeq
The hard-sphere potential is recovered in the limit $\ww\to\infty$, in which case $\beta=b/\sigma$, $\beta_0(\beta)=\text{min}(1,\beta)$, $\beta(\chi)=\cos (\chi/2)$, and $B(\gw,\chi)=\frac{1}{4}\sigma^2=\text{const}$. On the other hand, in the case of the Maxwell potential ($\ww=4$), the collision rate $\gw B(\gw,\chi)$ is independent of the relative speed $\gw$, namely
\beq
\gw B(\gw,\chi)=Q\mathcal{B}(\chi),\quad Q\equiv \sigma^2\sqrt{8\varphi_0/m}.
\label{2.7}
\eeq
Maxwell himself realized that if $\ww=4$ the integral in Eq.\ \eqref{2.5} can be expressed in terms of the complete elliptic integral of the first kind $K(z)$, namely
\beq
\chi(\beta)=\pi-2\frac{\beta}{(2+\beta^4)^{1/4}}K\left(\frac{1}{2}-\frac{\beta^2}{2(2+\beta^4)^{1/2}}\right).
\label{2.8}
\eeq
Obviously, the associated function $\mathcal{B}(\chi)$ for the IPL Maxwell potential becomes rather cumbersome and needs to be evaluated numerically. Nevertheless, it is sometimes convenient to depart from a strict adherence to the interaction potential $\varphi(r)$ by directly modeling the scattering angle dependence of the differential cross section \cite{E81}. In particular,  in the
variable soft-sphere (VSS) model  one has \cite{KM91,KM92}
\beq
\beta(\chi)= \cos^\alphak \frac{\chi}{2},\quad \mathcal{B}(\chi)=\frac{\alphak}{4}
\cos^{2(\alphak-1)} \frac{\chi}{2}.
\label{2.9}
\eeq
If one chooses $\alphak=1$, then the scattering is assumed to be
isotropic \cite{HN82} and one is dealing with the variable hard-sphere (VHS)
model \cite{B94}. However, this leads to a viscosity/self-diffusion ratio
different from that of the true IPL Maxwell interaction. To remedy this,
in the VSS model \cite{KM91} one takes $\alphak=2.13986$. In the context of \emph{inelastic} Maxwell models \cite{BNK03}, it is usual to take $\mathcal{B}(\chi)\propto|\widehat{\mathbf{g}}\cdot \widehat{\bm{\sigma}}|^{-1}=1/\sin(\chi/2)$.

\subsection{Moment equations}

Let $\psi(\mathbf{v})$ be a test velocity function. Multiplying both sides of the Boltzmann equation \eqref{2.1} by $\psi(\mathbf{v})$ and integrating over velocity one gets the balance equation
\beq
\frac{\partial \Psi}{\partial t}+\nabla\cdot \bm{\Phi}_\psi=\sigma_\psi^{(F)}+J_\psi,
\label{2.10}
\eeq
where
\beq
\Psi=\int d\mathbf{v}\, \psi(\mathbf{v})f(\mathbf{v})\equiv n\langle \psi\rangle
\label{2.11}
\eeq
is the local density of the quantity represented by $\psi(\mathbf{v})$,
\beq
\bm{\Phi}_\psi= \int d\mathbf{v}\, \mathbf{v}\psi(\mathbf{v})f(\mathbf{v})
\label{2.12}
\eeq
is the associated flux,
\beq
\sigma_\psi^{(F)}=\int d\mathbf{v}\, \frac{\partial \psi}{\partial\mathbf{v}}\cdot \left[\frac{\mathbf{F}}{m} f(\mathbf{v})\right]
\label{2.13}
\eeq
is a  source term due to the external force, and
\beqa
J_\psi&=&\int d\mathbf{v}\,  \psi(\mathbf{v})J[\mathbf{v}|f,f]\nonumber\\
&=&\frac{1}{4}\int d\mathbf{v}\int d\mathbf{v}_1\int d\Omega\,
\gw B(\gw,\chi) \left[\psi(\mathbf{v})+\psi(\mathbf{v}_1)-\psi(\mathbf{v}')-\psi(\mathbf{v}_1')\right]f(\mathbf{v})f(\mathbf{v}_1)
\label{2.14}
\eeqa
is the source term due to collisions.
The general balance equation \eqref{2.10}, which is usually  referred to as the weak form of the Boltzmann equation, is close to the approach followed by Maxwell in 1867 \cite{B03,TM80}.
We say that $\Psi$ is a moment  of order $\kk$ if $\psi(\mathbf{v})$ is a polynomial of degree $\kk$. In that case, if the external force $\mathbf{F}$ is independent of velocity, the source term $\sigma_\psi^{(F)}$ is a moment of order $\kk-1$. The hierarchical structure of the moment equations is in general due to the flux $\bm{\Phi}_\psi$ and the collisional moment $J_\psi$. The former is a moment of order $\kk+1$, while the latter is a bilinear combination of moments of any order because of the velocity dependence of the collision rate $\gw B(\gw,\chi)$. In order to get a closed set of equations some kind of approximate closure needs to be applied. In the Hilbert and Chapman--Enskog (CE) methods \cite{B81,CC70,S09} one focuses on the balance equations for the five conserved quantities, namely $\psi(\mathbf{v})=\left\{1,\mathbf{v},v^2\right\}$, so that $J_\psi=\left\{0,\mathbf{0},0\right\}$. Next, an expansion of the velocity distribution function in powers of the Knudsen number $\text{Kn}=\ell_{\text{mfp}}/\ell_h$ (defined as the ratio between the mean free path $\ell_{\text{mfp}}$ and the characteristic distance $\ell_h$ associated with the hydrodynamic gradients) provides the Navier--Stokes (NS) hydrodynamic equations and their sequels (Burnett, super-Burnett, \ldots). On a different vein, Grad proposed in 1949 \cite{G49,G63} to expand the distribution function $f$ in a complete set of orthogonal polynomials (essentially Hermite polynomials), the coefficients being the corresponding velocity moments. Next, this expansion is truncated by retaining terms up to a given order $\kk$, so the (orthogonal) moments of order higher than $\kk$ are neglected and one finally gets a {closed} set of moment equations.
In the usual 13-moment approximation, the expansion includes the density $n$, the three components of the  flow velocity $\mathbf{u}$,  the six  elements of the pressure tensor $\mathsf{P}$, and the three components of the  heat flux $\mathbf{q}$. The method can be augmented to twenty moments by including  the seven third-order moments apart from the heat flux. Variants of Grad's moment method have been developed in the last few years \cite{S05}, this special issue reflecting the current state of the art.

As said above, the collisional moment $J_\psi$ involves in general moments of every order. An important exception takes place in the case of  Maxwell models, where $\gw B(\gw,\chi)$ is independent of the relative speed $\gw$ [cf.\ Eq.\ \eqref{2.6} with $\ww=4$]. This implies that, if $\psi(\mathbf{v})$ is a polynomial of degree $\kk$, then $J_\psi$ becomes a bilinear combination of moments of order equal to or less than $\kk$ \cite{HN82,M89,TM80}. To be more precise, let us define the reduced orthogonal moments \cite{C90,G49,M89,RL77,TM80}
\begin{equation}
\psi_{\rr\ell \mu}(\mathbf{c})=N_{\rr\ell} c^{\ell}
L^{(\ell+\frac{1}{2})}_{\rr}(c^2)Y_{\ell}^\mu(\widehat{\mathbf{c}}),
\label{1.4:22}
\end{equation}
where
\beq
\mathbf{c}=\sqrt{\frac{m}{2k_BT}}{\bf V}
\label{1.4:23}
\eeq
is the peculiar
velocity normalized with respect to the  thermal velocity,
 $L^{(\ell+\frac{1}{2})}_{\rr}(c^2)$ are
generalized Laguerre polynomials \cite{AS72,GR80} (also known as  Sonine
polynomials),  $Y_{\ell}^\mu(\widehat{\mathbf{c}})$ are spherical harmonics, $\widehat{\mathbf{c}}=\mathbf{c}/c$ being the unit vector along the direction of $\mathbf{c}$, and
\begin{equation}
N_{\rr\ell}=\left[ 2 \pi^{3/2} \frac{\rr!}{\Gamma(\rr+\ell+\frac{3}{2})}
\right]^{1/2}
\label{1.4:24bis}
\end{equation}
are normalization constants, $\Gamma(x)$ being the gamma function. The polynomials $\{\psi_{\bm{\kk}}(\mathbf{c});\bm{\kk}\equiv (\rr\ell\mu)\}$ form a complete set of orthonormal functions with respect to the inner product $\langle F|G\rangle=\pi^{-3/2} \int d\mathbf{c}\, e^{-c^2}F^*(\mathbf{c})G(\mathbf{c})$. Let us denote as
\beq
\mathcal{M}_{\bm{\kk}}=\frac{1}{n}\int d\mathbf{v}\, \psi_{\bm{\kk}}(\mathbf{c}) f(\mathbf{v})
\label{2.15}
\eeq
the (reduced) moment of order $\kk=2\rr+\ell$ associated with the polynomial $\psi_{\bm{\kk}}(\mathbf{c})\equiv\psi_{\rr\ell\mu}(\mathbf{c})$. In the case of Maxwell models one has
\beqa
\mathcal{J}_{\bm{\kk}}&=&\frac{1}{n}\int d\mathbf{v}\, \psi_{\bm{\kk}}(\mathbf{c}) J[\mathbf{v}|f,f]\nn
&=&-\lambda_{\rr\ell}\mathcal{M}_{\bm{\kk}}+\sum_{\bm{\kk}',\bm{\kk}''}^\dagger C_{\bm{\kk}\bm{\kk}'\bm{\kk}''}\mathcal{M}_{\bm{\kk}'}\mathcal{M}_{\bm{\kk}''}.
\label{2.16}
\eeqa
The dagger in the summation means the constraints  $\kk'+\kk''=\kk$ and $2\leq \kk'\leq \kk-2$. Moreover, it is necessary that  $(\ell,\ell',\ell'')$ form a triangle (i.e.,  $|\ell'-\ell''|\leq\ell\leq\ell'+\ell''$) and $\mu'+\mu''=\mu$ \cite{M89}. The explicit expression of  the coefficients $C_{\bm{\kk}\bm{\kk}'\bm{\kk}''}=C_{\bm{\kk}\bm{\kk}''\bm{\kk}'}$ is rather involved and can be found in Ref.\ \cite{M89}. A computer-aided algorithm to generate the collisional moments for Maxwell models is described in Ref.\ \cite{ST02}.

\begin{table}[t]
\caption{Reduced eigenvalues $\lambdabar_{\rr\ell}\equiv \lambda_{\rr\ell}/\lambda_{11}$ for $2\rr+\ell\leq 6$}
\centering
\label{tab:0}       
\begin{tabular}{ccc}
\hline\noalign{\smallskip}
$\rr$ & $\ell$&$\lambdabar_{\rr\ell}$\\[3pt]
\tableheadseprule\noalign{\smallskip}
$0$&$0$&$0$\\
\hline\noalign{\smallskip}
$0$&$1$&$0$\\
\hline\noalign{\smallskip}
$1$&$0$&$0$\\
$0$&$2$&$\frac{3}{2}$\\
\hline\noalign{\smallskip}
$1$&$1$&$1$\\
$0$&$3$&$\frac{9}{4}$\\
\hline\noalign{\smallskip}
$2$&$0$&$1$\\
$1$&$2$&$\frac{7}{4}$\\
$0$&$4$&$\frac{7}{2}-\frac{35}{8}\frac{A_4}{A_2}$\\
\hline\noalign{\smallskip}
$2$&$1$&$\frac{3}{2}$\\
$1$&$3$&$\frac{11}{4}-\frac{5}{2}\frac{A_4}{A_2}$\\
$0$&$5$&$5-\frac{175}{16}\frac{A_4}{A_2}$\\
\hline\noalign{\smallskip}
$3$&$0$&$\frac{3}{2}$\\
$2$&$2$&$\frac{9}{4}-\frac{3}{2}\frac{A_4}{A_2}$\\
$1$&$4$&${4}-\frac{115}{16}\frac{A_4}{A_2}$\\
$0$&$6$&$\frac{27}{4}-\frac{357}{16}\frac{A_4}{A_2}+\frac{231}{16}\frac{A_6}{A_2}$\\
\noalign{\smallskip}\hline
\end{tabular}
\end{table}

Equation \eqref{2.16} shows that the polynomials $\psi_{\bm{\kk}}(\mathbf{c})$ are \emph{eigenfunctions} of the \emph{linearized} collision operator for Maxwell particles, the \emph{eigenvalues} $\lambda_{\rr\ell}$ being given by
\beq
\lambda_{\rr\ell}= 2\pi nQ \int_0^\pi d\chi\,\sin\chi
\mathcal{B}(\chi)\left[1+\delta_{r0}\delta_{\ell 0}-
\cos^{2\rr+\ell}\frac{\chi}{2}P_\ell\left(\cos\frac{\chi}{2}\right)-
\sin^{2\rr+\ell}\frac{\chi}{2}P_\ell\left(\sin\frac{\chi}{2}\right)\right],
\label{2}
\eeq
where $P_\ell(x)$ are Legendre
polynomials.
Irrespective of the precise angular dependence of
$\mathcal{B}(\chi)$,  the following relationships hold
\beq
\lambda_{\rr,1}=\lambda_{\rr+1,0},\quad (2\ell+1)\lambda_{\rr\ell}=(\ell+1)\lambda_{\rr-1,\ell+1}+\ell \lambda_{\rr,\ell-1}.
\label{3a}
\eeq
The eigenvalues can  be expressed as linear combinations of the numerical coefficients
\beqa
A_{2a}&=& 2\pi \int_0^\pi d\chi\,\sin\chi \mathcal{B}(\chi)
\cos^a\frac{\chi}{2} \sin^{a}\frac{\chi}{2}\nn
&=&  2\pi \int_0^\infty d\beta\,\beta
\cos^{2a}\frac{\chi(\beta)}{2} \sin^{2a}\frac{\chi(\beta)}{2}.
\label{7}
\eeqa
In particular, $\lambda_{11}=2nQA_{2}$. The reduced eigenvalues $\lambdabar_{\rr\ell}\equiv \lambda_{\rr\ell}/\lambda_{11}$ of order $2\rr+\ell\leq 6$ are listed in Table \ref{tab:0}.
For the IPL model the coefficients $A_a$ must be evaluated numerically. On the other hand, by assuming Eq.\ \eqref{2.9} one simply has
\beq
A_{2a}=\pi \alphak \text{B}(a+\alphak,a+1),
\label{2.18}
\eeq
where $\text{B}(x,y)=\Gamma(x)\Gamma(y)/\Gamma(x+y)$ is the Euler beta
function \cite{AS72,GR80}.
Table \ref{tab:1} gives the first few values of $A_{2a}$ for the IPL, VSS, and VHS models. It can be observed that the latter two models provide values quite close to the correct IPL ones, especially as $a$ increases.
A rather extensive table of the eigenvalues $\lambda_{\rr\ell}$ for the IPL Maxwell model can be found in Ref.\ \cite{AFP62}.

The most important eigenvalues are $\lambda_{02}$ and $\lambda_{11}$, which  provide the NS shear viscosity and thermal conductivity, namely
\beq
P_{ij}^{\text{NS}}=p\delta_{ij}-\eta_{\text{NS}}\left(\nabla_i u_j+\nabla_j u_i-\frac{2}{3}\nabla\cdot \mathbf{u}\delta_{ij}\right),\quad \eta_{\text{NS}}=\frac{p}{\lambda_{02}},
\label{2.19}
\eeq
\beq
\mathbf{q}^{\text{NS}}=-\kappa_{\text{NS}}\nabla T,\quad \kappa_{\text{NS}}=\frac{5}{2}\frac{k_Bp}{m\lambda_{11}}.
\label{2.20}
\eeq

\begin{table}[t]
\caption{Coefficients $A_{2a}$ ($a=1,\ldots 5$) for the IPL, VSS ($\alphak=2.13986$), and VHS ($\alphak=1$) models}
\centering
\label{tab:1}       
\begin{tabular}{cccc}
\hline\noalign{\smallskip}
$a$ & ILP & VSS&VHS  \\[3pt]
\tableheadseprule\noalign{\smallskip}
$1$&$0.685174$&$0.517177$&$0.523599$\\
$2$&$0.108109$&$0.102913$&$0.104720$\\
$3$&$0.0213745$&$0.0219923$&$0.0224399$\\
$4$&$0.00455921$&$0.00487876$&$0.00498666$\\
$5$&$0.00101072$&$0.00110750$&$ 0.00113333$\\
\noalign{\smallskip}\hline
\end{tabular}
\end{table}

\subsection{Solvable states}
Even in the case of Maxwell models the moment hierarchy \eqref{2.10} couples moments of order $\kk$ to moments of order $\kk+1$ through the flux term $\bm{\Phi}_\psi$. Therefore, the moment equations cannot be solved  in general unless an approximate closure is introduced. There exist, however, a few steady states where the hierarchy \eqref{2.10} for Maxwell molecules can be \emph{recursively} solved \cite{GS03,SG95b}. In those solutions one focuses on the \emph{bulk} of the system  (i.e., away from the boundary layers) and assumes that the velocity distribution function adopts a \emph{normal} form, i.e., it  depends on space only through the hydrodynamic fields ($n$, $\mathbf{u}$, and $T$) and their gradients.  On the other hand, it is not necessary to invoke that the  Knudsen number  $\text{Kn}=\ell_{\text{mfp}}/\ell_h$ (where, as said before, $\ell_{\text{mfp}}$ is the mean free path and $\ell_h$ is the characteristic distance associated with the hydrodynamic gradients) is small.

\begin{figure*}
\centering
  \includegraphics[width=0.85\textwidth]{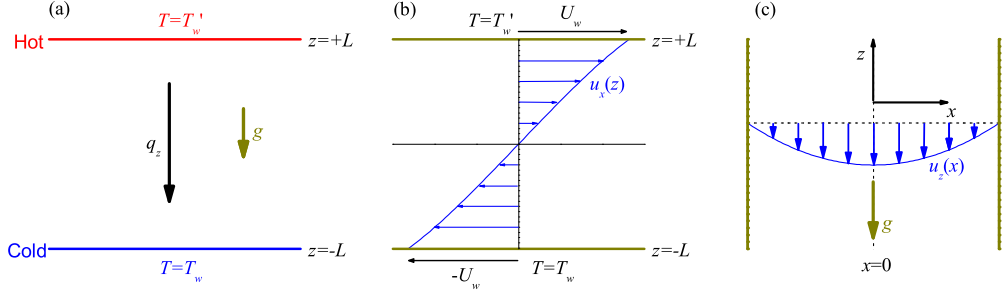}
\caption{Sketch of  the steady states considered in this paper: (a) Fourier flow without ($g=0$) or with ($g\neq 0$) gravity, (b) Couette flow, and (c) force-driven Poiseuille flow.}
\label{sketch1}       
\end{figure*}

The aim of the remainder of the paper is to review some of those solutions. All of them have the common features of a planar or channel geometry (gas enclosed between infinite parallel plates) and a one-dimensional spatial dependence along the direction orthogonal to the plates. Figure \ref{sketch1}  sketches the four steady states to be considered. In the planar Fourier flow (without gravity) the recursive solvability of the moment equations is tied to the fact that the (reduced) moments of order $\kk=2\rr+\ell$ are just polynomials in the Knudsen number (here associated with the thermal gradient) of degree $\kk-2$ and parity $\ell$. This simple polynomial dependence is broken down when a gravity field is added  but the problem is still solvable by a perturbation expansion in powers of  the gravity strength.  When the gas is sheared by moving plates (planar Couette flow) the dependence of the reduced moments on the
Knudsen number  associated with the thermal gradient is still polynomial, but with coefficients that depend on the (reduced) shear rate. In the case of the force-driven Poiseuille flow the nonequilibrium hydrodynamic profiles are induced by the presence of a longitudinal body force only. Again a perturbation expansion allows one to get those profiles, which exhibit interesting non-Newtonian features.

\section{Planar Fourier flow}
\label{sec3}

\subsection{Without gravity}
\label{sec3.1}
In the planar Fourier flow the gas is enclosed between two infinite parallel plates kept at different temperatures [see Fig.\ \ref{sketch1}(a)]. We assume that no external force is acting on the particles ($g=0$) and consider the steady state of the gas with gradients along the direction normal to the plates ($\nabla\to \widehat{\bm{z}} \partial_z $) and no flow velocity,
\beq
\mathbf{u}=\mathbf{0}.
\label{u=0}
\eeq
Under these conditions, the Boltzmann equation \eqref{2.1} becomes
\beq
v_z\frac{\partial}{\partial z} f(z,\mathbf{v})=J[\mathbf{v}|f,f]
\label{3.0}
\eeq
and the conservation laws of momentum and energy yield
\beq
{P_{zz}}=\text{const},
\label{4.4a}
\eeq
\beq
q_z=\text{const},
\label{4.5}
\eeq
respectively.

In 1979 Asmolov, Makashev, and Nosik \cite{AMN79} proved that an exact \emph{normal} solution  of Eq.\ \eqref{3.0}  for Maxwell molecules exists with
\begin{subequations}
\label{3.1}
\beq
p=nk_BT=\text{const},
\label{3.1a}
\eeq
\beq
 T(z)\frac{\partial T(z)}{\partial z}=\text{const}.
\label{3.1b}
\eeq
\end{subequations}
The original paper is rather  condensed and difficult to follow, so an independent derivation \cite{SBD87} is expounded here.

Since in this problem the only hydrodynamic gradient is the thermal one, i.e., $\partial_z T$, the obvious choice of hydrodynamic length is $\ell_{T}=(\partial_z\ln T)^{-1}$. As for the mean free path one can take $\ell_{\text{mfp}}=\sqrt{2k_BT/m}/\lambda_{11}$. Both quantities are local and their ratio defines the relevant \emph{local} Knudsen number of the problem, namely
\beq
\epsilon(z)=\frac{\sqrt{2k_BT(z)/m}}{\lambda_{11}(z)}\frac{\partial\ln T(z)}{\partial z}.
\label{3.2}
\eeq
Note that $\lambda_{11}(z)\propto n(z)\propto 1/T(z)$, so that $\sqrt{T(z)}\epsilon(z)=\text{const}$, where use has been made of Eqs.\ \eqref{3.1}. Here it is assumed that the separation $2L$ between the plates is large enough as compared with the mean free path to identify a bulk region where the normal solution applies. Such a solution can be nondimensionalized in the form
\beq
\phi(\mathbf{c};\epsilon)=\frac{1}{n(z)}\left[\frac{2k_BT(z)}{m}\right]^{3/2}f(z,\mathbf{v}),
\label{3.3}
\eeq
where $\mathbf{c}$ is defined by Eq.\ \eqref{1.4:23}.
All the spatial dependence of $\phi$ is contained in its dependence on $\mathbf{c}$ and $\epsilon$. Thus
\beq
\frac{\partial f}{\partial z}=\frac{\partial T}{\partial z}\frac{\partial f}{\partial T}
,
\eeq
where
\beq
\frac{\partial f}{\partial T}
=n \left(\frac{2k_BT}{m}\right)^{-3/2}\left(-\frac{5}{2} T^{-1}\phi+\frac{\partial \mathbf{c}}{\partial T}\cdot \frac{\partial \phi}{\partial \mathbf{c}}
+\frac{\partial \epsilon}{\partial T} \frac{\partial \phi}{\partial \epsilon}\right).
\label{3.3.n}
\eeq
Taking into account that $\partial \mathbf{c}/\partial T=-\frac{1}{2}T^{-1}\mathbf{c}$ and  $\partial \epsilon/\partial T=-\frac{1}{2}T^{-1}\epsilon$, one finally gets
\beq
\frac{\partial}{\partial z}
f(z,\mathbf{v})=-n\left(\frac{m}{2k_BT}\right)^2\lambda_{11}\frac{\epsilon}{2}\left(2+
\frac{\partial}{\partial\mathbf{c}}\cdot\mathbf{c}+\epsilon\frac{\partial}{\partial
\epsilon}\right)\phi(\mathbf{c};\epsilon).
\label{2.4.n}
\eeq
Consequently,  the Boltzmann equation (\ref{3.0}) becomes
\beqa
-\frac{\epsilon}{2}c_z\left(2+
\frac{\partial}{\partial\mathbf{c}}\cdot\mathbf{c}+\epsilon\frac{\partial}{\partial
\epsilon}\right)\phi(\mathbf{c};\epsilon)&=& \frac{n}{\lambda_{11}} Q\int
d\mathbf{c}_1\int
d\Omega\,\mathcal{B}(\chi)\left[\phi(\mathbf{c}_1';\epsilon)\phi(\mathbf{c}';\epsilon)-\phi(\mathbf{c}_1;\epsilon)\phi(\mathbf{c};\epsilon)\right]\nn
&\equiv& \overline{J}[\mathbf{c}|\phi(\epsilon),\phi(\epsilon)].
\label{2.5.n}
\eeqa

The orthogonal moments of $\phi(\mathbf{c};\epsilon)$ are
\beq
\mathcal{M}_{\rr\ell}(\epsilon)=\int d\mathbf{c}\,\psi_{\rr\ell}(\mathbf{c})\phi(\mathbf{c};\epsilon),
\label{3.4}
\eeq
so that
\beq
\phi(\mathbf{c};\epsilon)=\pi^{-3/2} e^{-c^2}\sum_{\rr=0}^\infty\sum_{\ell=0}^\infty \mathcal{M}_{\rr\ell}(\epsilon)\psi_{\rr\ell}(\mathbf{c}).
\label{3.5}
\eeq
Here,
\beq
\psi_{\rr\ell}(\mathbf{c})\equiv \psi_{\rr\ell 0}(\mathbf{c})=N_{\rr\ell}\sqrt{\frac{2\ell+1}{4\pi}} L_{\rr}^{(\ell+\frac{1}{2})}(c^2)c^\ell P_\ell(c_z/c).
\eeq
The moments $\{\mathcal{M}_{\rr\ell}\}$ are a subclass of the moments defined by Eq.\ \eqref{2.15}. {}From the definition of temperature and density, and the fact that the flow velocity vanishes, it follows that
\beq
\mathcal{M}_{00}=1, \quad\mathcal{M}_{10}=\mathcal{M}_{01}=0.
\label{3.12}
\eeq
Taking into account the recurrence relations of the Laguerre and Legendre polynomials \cite{AS72,GR80} it is easy to prove that
\beq
\mathbf{c}\cdot\frac{\partial}{\partial\mathbf{c}}\psi_{\rr\ell}(\mathbf{c})=(2\rr+\ell)\psi_{\rr\ell}(\mathbf{c})-2\sqrt{\rr\left(\rr+\ell
+\frac{1}{2}\right)}\psi_{\rr-1,\ell}(\mathbf{c}),
\eeq
\beq
c_z\psi_{\rr\ell}(\mathbf{c})=\gammak_{\rr\ell}\psi_{\rr,\ell+1}(\mathbf{c})-\omegak_{\rr+1,\ell-1}\psi_{\rr+1,\ell-1}(\mathbf{c})
+\gammak_{\rr,\ell-1}\psi_{\rr,\ell-1}(\mathbf{c})-\omegak_{\rr\ell}\psi_{\rr-1,\ell+1}(\mathbf{c}),
\label{3.6}
\eeq
where
\beq
\gammak_{\rr\ell}\equiv (\ell+1)\left[\frac{\rr+\ell+\frac{3}{2}}{(2\ell+1)(2\ell+3)}\right]^{1/2},\quad \omegak_{\rr\ell}\equiv (\ell+1)\left[\frac{\rr}{(2\ell+1)(2\ell+3)}\right]^{1/2}.
\eeq

Multiplying both sides of Eq.\ \eqref{2.5.n} by
$\psi_{\rr\ell}(\mathbf{c})$ and integrating over $\mathbf{c}$ one gets
\beqa
&&\frac{\epsilon}{2}\left(2\rr+\ell-1-\epsilon\frac{\partial}{\partial\epsilon}\right)\left(\gammak_{\rr\ell}\mathcal{M}_{\rr,\ell+1}-
\omegak_{\rr+1,\ell-1}\mathcal{M}_{\rr+1,\ell-1}
+\gammak_{\rr,\ell-1}\mathcal{M}_{\rr,\ell-1}-\omegak_{\rr\ell}\mathcal{M}_{\rr-1,\ell+1}\right)\nn
&&-\epsilon\sqrt{\rr\left(\rr+\ell+\frac{1}{2}\right)}\left(\gammak_{\rr-1,\ell}\mathcal{M}_{\rr-1,\ell+1}-
\omegak_{\rr,\ell-1}\mathcal{M}_{\rr,\ell-1}
+\gammak_{\rr-1,\ell-1}\mathcal{M}_{\rr-1,\ell-1}-\omegak_{\rr-1,\ell}\mathcal{M}_{\rr-2,\ell+1}\right)\nn
&=&-\lambdabar_{\rr\ell}\mathcal{M}_{\rr\ell}+\frac{1}{\lambda_{11}}\sum_{\rr'\ell'\rr''\ell''}^\dagger C_{\rr\ell \rr'\ell' \rr''\ell''}\mathcal{M}_{\rr'\ell'}\mathcal{M}_{\rr''\ell''}.
\label{3.7}
\eeqa
The choice of the orthogonal moments $\{\mathcal{M}_{\rr\ell}\}$ simplifies the structure of the collisional terms but, on the other hand, complicates the convective terms. Alternatively, one can use the non-orthogonal moments
\beq
\overline{M}_{\rr\ell}(\epsilon)=\int d\mathbf{c}\,c^{2\rr}c_z^\ell\phi(\mathbf{c};\epsilon).
\label{3.8}
\eeq
In that case, one gets
\beqa
\frac{\epsilon}{2}\left(2\rr+\ell-1-\epsilon\frac{\partial}{\partial
\epsilon}\right)\overline{M}_{\rr,\ell+1}(\epsilon)&=&\int
d\mathbf{c}\,c^{2\rr}c_z^\ell \overline{J}[\mathbf{c}|\phi(\epsilon),\phi(\epsilon)]\nn
&\equiv&\overline{J}_{\rr\ell}(\epsilon)
\label{3.7bis}
\eeqa
from Eq.\ \eqref{2.5.n}. Obviously, $\overline{J}_{\rr\ell}$ is a bilinear combination of moments $\overline{M}_{\rr'\ell'}$ of order equal to or smaller than $\kk=2\rr+\ell$.

Equations \eqref{3.7} and \eqref{3.7bis} reveal the hierarchical character of the moment equations: moments of order $\kk$ are coupled to moments of lower order but also to moments of order $\kk+1$. However, the hierarchy can be solved by following a recursive scheme. Of course, the moments of zeroth and first order, as well as one of the two moments of second order are known by definition [cf.\ Eq.\ \eqref{3.12}]. The key point is that the moments of order $\kk\geq 2$ are \emph{polynomials} in $\epsilon$ of degree $\kk-2$ and parity $\ell$:
\beq
\mathcal{M}_{\rr\ell}(\epsilon)=\sum_{j=0}^{2\rr+\ell-2}{\mu}_j^{(\rr\ell)}\epsilon^j,\quad \mu_j^{(\rr\ell)}=0\text{ if }j+\ell=\text{odd},
\label{3.9}
\eeq
where $\mu_j^{(\rr\ell)}$ are pure numbers to be determined.
Let us see that Eq.\ \eqref{3.9} is consistent with Eq.\ \eqref{3.7}. First note that all the moments on the left-hand side of Eq.\ \eqref{3.7} have a parity different from that of $\mathcal{M}_{\rr\ell}$ but the parity is restored when multiplying  by $\epsilon$ or applying the operator $\epsilon^2\partial/\partial\epsilon$. Similarly, the condition $2\rr'+\ell'+2\rr''+\ell''=2\rr+\ell$ assures that the product $\mathcal{M}_{\rr'\ell'}\mathcal{M}_{\rr''\ell''}$ is a polynomial of the same degree and parity as those of $\mathcal{M}_{\rr\ell}$.
Next, although the moments $\mathcal{M}_{\rr,\ell+1}$ and $\mathcal{M}_{\rr+1,\ell-1}$ are polynomials of degree $2\rr+\ell-1$, the action of the operator $\epsilon(2\rr+\ell-1-\epsilon\partial/\partial\epsilon)$ transforms those polynomials into polynomials of degree $2\rr+\ell-2$.

\begin{figure*}
\centering
  \includegraphics[width=0.5\textwidth]{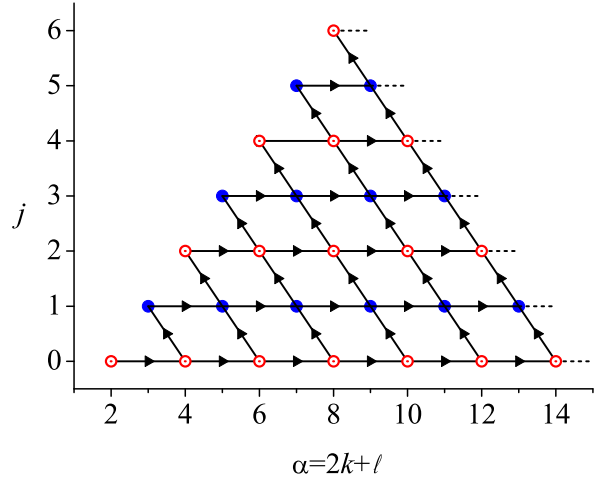}
\caption{Planar Fourier flow. Sketch of the sequence followed in the recursive determination of the numerical coefficients $\mu_j^{(\rr\ell)}$. All the coefficients of the same order $\kk=2\rr+\ell$ are represented by a common circle.
The coefficients with $j<\kk/3$ (represented by circles lying below the dash-dot line) are equal to zero. So are the coefficients with $j<\ell$.}
\label{arrows}       
\end{figure*}

In order to complete the proof that \eqref{3.9} provides a solution to the moment equations \eqref{3.7}, a recursive scheme must be devised to get the numerical coefficients $\mu_j^{(\rr\ell)}$. First, notice that the left-hand side of Eq.\ \eqref{3.7} contains at least the first power of $\epsilon$. So, if $\ell=\text{even}$ one easily gets $\mu_0^{(\rr\ell)}$, provided that $\mu_0^{(\rr'\ell')}$ is known for $2\rr'+\ell'\leq 2\rr+\ell-2$. Next, if $\ell=\text{odd}$, the coefficient $\mu_1^{(\rr\ell)}$ is determined from the previous knowledge of $\mu_0^{(\rr'\ell')}$ for $2\rr'+\ell'\leq 2\rr+\ell+1$ and of $\mu_1^{(\rr'\ell')}$ for $2\rr'+\ell'\leq 2\rr+\ell-2$. Again, if $\ell=\text{even}$ and we know $\mu_0^{(\rr'\ell')}$ and $\mu_2^{(\rr'\ell')}$ for $2\rr'+\ell'\leq 2\rr+\ell-2$, as well as $\mu_1^{(\rr'\ell')}$ for $2\rr'+\ell'\leq 2\rr+\ell+1$, we can get $\mu_2^{(\rr\ell)}$, and so on. As a starting point (moments of zeroth degree) we have $\mathcal{M}_{10}=0$ and $\mathcal{M}_{02}=0$,
the latter being a consequence of $\mathcal{M}_{01}=0$. The recursive scheme is sketched in Fig.\ \ref{arrows}, where the arrows indicate the sequence followed in the determination of $\mu_j^{(\rr\ell)}$. The open (closed) circles represent the coefficients associated with even (odd) $j$ and $\ell$. Notice that the number of coefficients needed to determine a given moment $\mathcal{M}_{\rr\ell}$ is finite. In fact, the coefficients represented in Fig.\ \ref{arrows} are the ones involved in the evaluation of $\mathcal{M}_{\rr\ell}$ for $\rr+2\ell\leq 8$. In practice,  the number of coefficients actually needed is smaller since
\beq
\mu_j^{(\rr\ell)}=0 \text{ if } j<\text{max}\left(\frac{2\rr+\ell}{3},\ell\right).
\label{3.10}
\eeq
To check the above property, note that Eq.\ \eqref{3.7} implies that
\beqa
\mu_j^{(\rr\ell)}&=&\text{L.C.}\left\{\mu_{j-1}^{(\rr,\ell+1)},\mu_{j-1}^{(\rr+1,\ell-1)},\mu_{j-1}^{(\rr-1,\ell+1)},\mu_{j-1}^{(\rr,\ell-1)},
\mu_{j-1}^{(\rr-2,\ell+1)},\mu_{j-1}^{(\rr-1,\ell-1)}\right\}\nn
&&+\text{L.C.}\left\{\sum_{j'=0}^j\mu_{j'}^{(\rr'\ell')}\mu_{j-j'}^{(\rr''\ell'')}\right\}_{2\rr'+\ell'+2\rr''+\ell''=2\rr+\ell, \; |\ell'-\ell''|\leq \ell\leq \ell'+\ell''},
\label{3.11}
\eeqa
where L.C. stands for ``linear combination of''. If $\mu_j^{(\rr\ell)}=0$ for $j<\frac{1}{3}(2\rr+\ell)$ then the first term on the right-hand side of Eq.\ \eqref{3.11} vanishes because  $j-1<\frac{1}{3}(2\rr+\ell-3)<\frac{1}{3}(2\rr+\ell-1)<\frac{1}{3}(2\rr+\ell+1)$; the second term also vanishes because either $j'<\frac{1}{3}(2\rr'+\ell')$ or $j-j'<\frac{1}{3}(2\rr''+\ell'')$. Similarly, if $\mu_j^{(\rr\ell)}=0$ for $j<\ell$ then both terms on the right-hand side  vanish because  $j-1<\ell-1<\ell+1$ and either $j'<\ell'$ or $j-j'<\ell''$, respectively.
The property $\mu_j^{(\rr\ell)}=0$ for $j<\ell$ yields
\beq
\mathcal{M}_{0\ell}=0,\quad \ell\geq 1.
\label{3.13}
\eeq


\begin{table}[t]
\caption{Planar Fourier flow. Orthogonal moments $\mathcal{M}_{\rr\ell}(\epsilon)$ and non-orthogonal moments $\overline{M}_{\rr\ell}(\epsilon)$ for $2\rr+\ell\leq 5$ and for $(\rr,\ell)=(3,1)$, $(4,1)$, and $(5,1)$. In the latter cases the numerical coefficients correspond to the IPL model. Those numerical values are, however, very similar to those corresponding to the VHS and VSS models \protect\cite{GTRTS06}}
\centering
\label{tab:2}       
\begin{tabular}{cccc}
\hline\noalign{\smallskip}
$\rr$ & $\ell$&$\mathcal{M}_{\rr\ell}(\epsilon)$&$\overline{M}_{\rr\ell}(\epsilon)$\\[3pt]
\tableheadseprule\noalign{\smallskip}
$0$&$0$&$1$&$1$\\
\hline\noalign{\smallskip}
$0$&$1$&$0$&$0$\\
\hline\noalign{\smallskip}
$1$&$0$&$0$&$\frac{3}{2}$\\
$0$&$2$&$0$&$\frac{1}{2}$\\
\hline\noalign{\smallskip}
$1$&$1$&$\frac{\sqrt{5}}{2}\epsilon$&$-\frac{5}{4}\epsilon$\\
$0$&$3$&$0$&$-\frac{3}{4}\epsilon$\\
\hline\noalign{\smallskip}
$2$&$0$&$\frac{7}{2}\sqrt{\frac{5}{6}}\epsilon^2$&$\frac{15}{4}+\frac{35}{4}\epsilon^2$\\
$1$&$2$&$-4\sqrt{\frac{2}{21}}\epsilon^2$&$\frac{5}{4}+\frac{17}{4}\epsilon^2$\\
$0$&$4$&$0$&$\frac{3}{4}+\frac{81}{28}\epsilon^2$\\
\hline\noalign{\smallskip}
$2$&$1$&$-\frac{17(225 - 122 A_4/A_2)}{18\sqrt{35}(3-2A_4/A_2)}\epsilon^3$&$-\frac{35}{4}-\frac{17(225 - 122 A_4/A_2)}{36(3-2A_4/A_2)}\epsilon^3$\\
$1$&$3$&$\sqrt{\frac{3}{5}}\frac{4(100-27 A_4/A_2)}{7 (11 - 10 A_4/A_2) (3 - 2 A_4/A_2)}\epsilon^3$&$-\frac{21}{4}\epsilon- \frac{63225 - 86656 A_4/A_2 + 29036 (A_4/A_2)^2}{84 (11 - 10 A_4/A_2) (3 - 2 A_4/A_2)}\epsilon^3$\\
$0$&$5$&$0$&$-\frac{15}{4}\epsilon- \frac{5(3125 - 2074 A_4/A_2) }{84 (11 - 10 A_4/A_2)}\epsilon^3$\\
\hline\noalign{\smallskip}
$3$&$1$&$21.86 \epsilon^3 + 402.2 \epsilon^5$&$-\frac{945}{16} \epsilon - 726.2 \epsilon^3 -
 4.371 \times 10^3\epsilon^5$\\
$4$&$1$&$-9.250 \epsilon^3 - 1.471\times 10^3 \epsilon^5 - 2.816\times 10^4\epsilon^7$&$-\frac{3465}{8} \epsilon - 1.107\times 10^4 \epsilon^3 -
 1.712\times 10^5 \epsilon^5 - 1.436\times 10^6 \epsilon^7$\\
$5$&$1$&$2.038\times 10^3 \epsilon^5 + 1.592\times 10^5 \epsilon^7 + 3.548\times10^6 \epsilon^9$&$-\frac{225\,225}{64} \epsilon - \frac{20\,286\,875}{128} \epsilon^3 -
 4.593\times 10^6 \epsilon^5 -
 9.292\times 10^7 \epsilon^7 -
 1.031\times 10^9 \epsilon^9$\\
\noalign{\smallskip}\hline
\end{tabular}
\end{table}

Since the non-orthogonal moments $\overline{M}_{\rr\ell}$ are linear combinations of the orthogonal moments $\mathcal{M}_{\rr'\ell'}$ with $2\rr'+\ell'\leq 2\rr+\ell$ and $\ell+\ell'=\text{even}$, the polynomial form \eqref{3.9}  also holds for $\overline{M}_{\rr\ell}$:
\beq
\overline{M}_{\rr\ell}(\epsilon)=\sum_{j=0}^{2\rr+\ell-2}\overline{m}_j^{(\rr\ell)}\epsilon^j,\quad \overline{m}_j^{(\rr\ell)}=0\text{ if }j+\ell=\text{odd}.
\label{3.14}
\eeq
The explicit expressions for the moments of order smaller than or equal to five are given in Table \ref{tab:2}. It also includes the moments of order seven, nine, and eleven corresponding to $\ell=1$ and $\rr=3$, $5$, and $7$, respectively. The result $\mathcal{M}_{02}=0$ or, equivalently, $\overline{M}_{02}=\frac{1}{2}$ imply that
\beq
P_{zz}=p,
\label{NS2}
\eeq
in agreement with Newton's law \eqref{2.19}. The most noteworthy outcome is the linear relationship $\mathcal{M}_{11}(\epsilon)=\frac{\sqrt{5}}{2}\epsilon$ or, equivalently, $\overline{M}_{11}(\epsilon)=-\frac{{5}}{4}\epsilon$. This means that Fourier's law \eqref{2.20} is \emph{exactly} verified, i.e.,
\beq
q_z=q_z^{\text{NS}},
\label{NS1}
\eeq
no matter how large the temperature gradient is. Likewise,  $\mathcal{M}_{03}(\epsilon)=0$ or $\overline{M}_{03}(\epsilon)=-\frac{{3}}{4}\epsilon$ imply that
\beq
 \langle v_z^3\rangle=\frac{3}{5}
\langle v^2v_z\rangle.
\label{vz^3}
\eeq
The nonlinear character of the solution, however, appears through moments of fourth order and higher. Although those moments are polynomials, the velocity distribution function \eqref{3.5} involves all the powers in the Knudsen number $\epsilon$.

\begin{figure*}
\centering
  \includegraphics[width=0.7\textwidth]{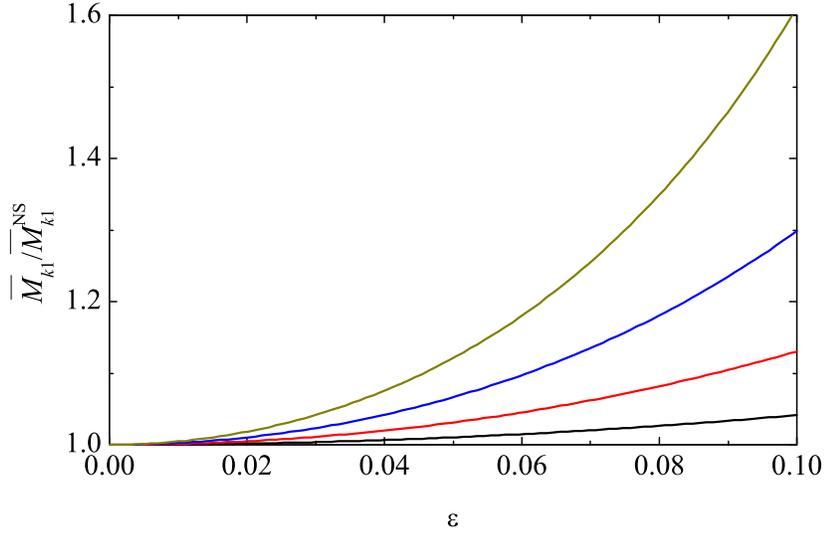}
\caption{Planar Fourier flow. Plot of the ratio $\overline{M}_{\rr 1}(\epsilon)/\overline{M}_{\rr 1}^{\text{NS}}(\epsilon)$ for, from bottom to top, $\rr=2$, 3, 4, and 5.}
\label{Mr1}       
\end{figure*}

Figure \ref{Mr1} shows the Knudsen number dependence of the moments $\overline{M}_{\rr 1}(\epsilon)$ ($\rr=2$--$5$) relative to their NS values  $\overline{M}_{\rr 1}^{\text{NS}}(\epsilon)=\overline{m}_1^{(\rr 1)}\epsilon$. It is quite apparent that, as  $\epsilon$ increases, important deviations from the NS predictions exist, even for low values of $\rr$. This is a consequence of the rapid growth of the nonlinear coefficients of $\mathcal{M}_{\rr 1}(\epsilon)$ and $\overline{M}_{\rr 1}(\epsilon)$ as $\rr$ increases. This property also holds in the exact solution \cite{KDSB89a,SBG86,SBKD89} of the Bhatnagar--Gross--Krook (BGK) model kinetic equation \cite{BGK54,W54} for the Fourier flow, where it is proven that it is closely related to the \emph{divergence} of the CE expansion \cite{SBKD89}. Whether or not this divergent character of the CE expansion also holds for the solution of the true Boltzmann equation for Maxwell molecules cannot be proven at this stage but it seems sensible to conjecture that the answer is affirmative.

A comparison between the analytical expressions of $\mathcal{M}_{\rr 1}(\epsilon)/\mathcal{M}_{11}(\epsilon)$ for $\rr=2$--$5$ and the numerical values obtained from the DSMC method for $\epsilon\lesssim 0.08$ shows an excellent agreement \cite{GTRTS06,GTRTS07}, thus validating both the practical applicability of the theoretical results in the bulk region and the reliability of the simulation method under highly nonequilibrium conditions.
Although the validity of Fourier's law for large thermal gradients is in principle restricted to Maxwell models, it turns out that its practical applicability extends to other potentials, such as that of hard spheres \cite{MASG94}.

Before closing this Section, it is worthwhile addressing the spatial dependence of the \emph{dimensional} moments
\beqa
M_{\rr\ell}(z)&=&\int d\mathbf{v}\, v^{2\rr}v_z^\ell  f(z,\mathbf{v})\nn
&=& n(z) \left[\frac{2k_BT(z)}{m}\right]^{\rr+\ell/2}\overline{M}_{\rr\ell}(\epsilon(z)).
\label{3.15}
\eeqa
Taking into account Eq.\ \eqref{3.14} and the fact that $\sqrt{T(z)}\epsilon(z)=\sqrt{T_0}\epsilon_0=\text{const}$ (where $T_0$ and $\epsilon_0$ are the temperature and the Knudsen number evaluated at $z=0$, respectively), one can easily get
\beq
M_{\rr\ell}(z)=\frac{2p}{m}\left(\epsilon_0^2\frac{2k_BT_0}{m}\right)^{\rr+\ell/2-1}\sum_{j=0}^{\rr-1+[\ell/2]}
\overline{m}_{2\rr+\ell-2-2j}^{(\rr\ell)}\left[\frac{T(z)}{\epsilon_0^2T_0}\right]^j,
\label{3.16}
\eeq
where $[\ell/2]$ is the integer part of $\ell/2$ and it is understood that $2\rr+\ell\geq 2$. Equation \eqref{3.16} shows that $M_{\rr\ell}(z)$ is just a polynomial in $T(z)$ of degree $\rr-1+[\ell/2]$. As for the spatial dependence of temperature, Eq.\ \eqref{3.1b} implies that
\beq
T(z)=T_0\sqrt{1+\frac{2\epsilon_0}{\ell_{\text{mfp}}(0)}z},
\label{3.17}
\eeq
where $\ell_\text{mfp}(0)\propto T_0^{3/2}/p$ is the local mean free path at $z=0$.
It is convenient to define a \emph{scaled} spatial variable $s$ as
\beq
s=\int_0^z dz'\, \lambda_{11}(z').
\label{3.18}
\eeq
Since $\lambda_{11}(z)\propto p/T(z)$, one has
\beq
s=\frac{\sqrt{2k_BT_0/m}}{\epsilon_0}\left[\sqrt{1+\frac{2\epsilon_0}{\ell_{\text{mfp}}(0)}z}-1\right].
\label{3.19}
\eeq
In terms of this new variable the temperature profile \eqref{3.17} becomes linear,
\beq
T(s)=T_0\left(1+\frac{\epsilon_0}{\sqrt{2k_BT_0/m}}s\right).
\label{3.20}
\eeq
Likewise, according to Eq.\ \eqref{3.16}, the moment $M_{\rr\ell}(s)$ is a polynomial in $s$ of degree $\rr-1+[\ell/2]$.
The exact solution of the BGK model for any interaction  potential \cite{KDSB89a,SBG86,SBKD89} keeps the linear and polynomial forms of $T(s)$ and  $M_{\rr\ell}(s)$, respectively, as functions of $s$. The only influence of the potential appears through the temperature dependence of the collision frequency (which plays the role of $\lambda_{11}$), so that when applying Eq.\ \eqref{3.18} one does not get Eqs.\ \eqref{3.17} and \eqref{3.19}, except for Maxwell molecules. The solution of the Boltzmann equation for Maxwell molecules described here can be extended to the case of gaseous mixtures \cite{N81}.

\subsection{With gravity}
Let us suppose that the planar Fourier flow is perturbed by a constant body force $\mathbf{F}=-mg \widehat{\mathbf{z}}$ orthogonal to the plates and directed downwards (e.g., gravity). This situation is sketched in Fig.\ \ref{sketch1}(a) with $g\neq 0$ and has the same geometry as the Rayleigh--B\'enard problem. In fact, we can define a \emph{microscopic} local Rayleigh number as
\beq
\gamma\equiv  \frac{g}{\lambda_{11}^2}\frac{\partial\ln T}{\partial z}=\frac{g\epsilon}{\lambda_{11}\sqrt{2k_BT/m}},
\label{4.1}
\eeq
where in the last step use has been made of Eq.\ \eqref{3.2}.
The dimensionless parameters $\gamma$ and $\epsilon$  characterize the normal state of the system. Note that $\gamma=0$ implies that either $g=0$ but $\epsilon\neq 0$ (Fourier flow without gravity) or $\epsilon=0$ but $g\neq 0$ (equilibrium state with a pressure profile given by the barometric formula).

The conventional Rayleigh number is $\text{Ra}\sim |\gamma|\left({L}/{\ell_\text{mfp}}\right)^4$. The situation depicted in Fig.\ \ref{sketch1}(a) corresponds to a gas heated from above ($\partial_zT>0\Rightarrow\gamma>0$). If the gas is heated from below then $\partial_zT<0$ and $\gamma<0$. We assume that either $\gamma>0$ or $\gamma<0$ but $\text{Ra}\lesssim 1700$, so that the gas  at rest is stable and there is no convection ($\mathbf{u}=\mathbf{0}$). On the other hand, rarefied gases can present a Rayleigh--B\'enard instability under appropriate conditions \cite{CS92,SAS97}.

The stationary Boltzmann equation for the problem at hand reads
\beq
v_z\frac{\partial}{\partial z} f(z,\mathbf{v})-g\frac{\partial}{\partial v_z}f(z,\mathbf{v})=J[\mathbf{v}|f,f].
\label{4.2}
\eeq
In this case the moment hierarchy \eqref{2.10} becomes
\beq
\frac{\partial}{\partial z}M_{\rr,\ell+1}(z)+g\left[2\rr M_{\rr-1,\ell+1}(z)+\ell M_{\rr,\ell-1}(z)\right] =J_{\rr\ell}(z),
\label{4.3}
\eeq
where the moments $M_{\rr\ell}$ are defined by the first equality of Eq.\ \eqref{3.15} and $J_{\rr\ell}$ are the corresponding collisional moments.
In particular, conservation of momentum and energy imply that $\partial_z M_{02}=-gn$ and $\partial_z M_{11}=0$, respectively. In other words, one has
\beq
\frac{\partial P_{zz}(z)}{\partial z}=-g\rho(z),
\label{4.4}
\eeq
where $\rho=mn$ is the mass density, plus Eq.\ \eqref{4.5}.
The presence of the  terms proportional to $g$ in Eqs.\ \eqref{4.2} and \eqref{4.3} complicates the problem significantly, thus preventing an exact solution (for general $\epsilon$ and $\gamma$), even in the case of Maxwell molecules. On the other hand, if $|\gamma|\ll 1$ one can treat it as a small parameter and carry out a perturbation expansion of the form
\beq
f(z,\mathbf{v})=f^{(0)}(z,\mathbf{v})+f^{(1)}(z,\mathbf{v})\gamma+f^{(2)}(z,\mathbf{v})\gamma^2+\cdots,
\label{4.6}
\eeq
\beq
M_{\rr\ell}(z)=M_{\rr\ell}^{(0)}(z)+M_{\rr\ell}^{(1)}(z)\gamma+M_{\rr\ell}^{(2)}(z)\gamma^2+\cdots.
\label{4.7}
\eeq
Here the superscript $(0)$ denotes the reference state without gravity, i.e., the one analyzed in Sec.\ \ref{sec3.1}. When the expansion \eqref{4.7} is inserted into Eq.\ \eqref{4.3} one  gets a recursive scheme allowing one to obtain, in principle, $\{M_{\rr\ell}^{(j+1)}\}$ from the previous knowledge of $\{M_{\rr\ell}^{(j)}\}$. The technical details can be found in Ref.\ \cite{TGS97} and here only the relevant final results will be presented. First, it turns out that the temperature field $T^{(j)}$ of order $\gamma^j$ is a polynomial of degree $j+1$ in the scaled variable $s$ defined by Eq.\ \eqref{3.18}. This generalizes the linear relationship \eqref{3.20} found for $T^{(0)}$ (i.e., in the absence of gravity). Analogously, $M_{\rr\ell}^{(j)}$ for $2\rr+\ell\geq 2$ is a polynomial in $s$ of degree $\rr+j-1+[\ell/2]$. In order to fully determine $M_{\rr\ell}^{(j)}$ it is necessary to make use of the collisional moments $J_{\rr'\ell'}$ with $2\rr'+\ell'\leq 2(2\rr+\ell+j-1)$.
The main results  are \cite{TGS97}
\beq
\frac{\partial}{\partial z} T\frac{\partial T}{\partial z}=-\frac{104}{5}\frac{m\lambda_{11}^2T }{k_B}\gamma^2
+\mathcal{O}(\gamma^3),
\label{4.9}
\eeq
\beq
\frac{P_{zz}-p}{p}=\frac{128}{45}\gamma^2+\mathcal{O}(\gamma^3),
\label{4.10}
\eeq
\beq
q_z=q_z^{\text{NS}}
\left[1+\frac{46}{5}\gamma+
\mathcal{O}(\gamma^2)
\right],
\label{4.11}
\eeq
\beq
{\langle v_z^3\rangle}=\frac{3}{5}{\langle v^2v_z\rangle}\left[1+\frac{64}{105}\gamma+
\mathcal{O}(\gamma^2)
\right].
\label{4.12}
\eeq
Equations \eqref{4.9}--\eqref{4.12} account for the first corrections to Eqs.\ \eqref{3.1b}, \eqref{NS2}, \eqref{NS1}, and \eqref{vz^3}, respectively, due to the presence of gravity.
In addition, given Eqs.\ \eqref{4.4} and \eqref{4.10}, one has
${\partial_z p}=-\rho g +\mathcal{O}(\gamma^3)$. According to Eq.\ \eqref{4.11} the presence of gravity produces an enhancement of the heat flux with respect to its NS value when heating from above ($\gamma>0$). The opposite effect occurs when heating from below ($\gamma<0$).

The above theoretical predictions for Maxwell molecules were seen to agree at a semi-quantitative level with DSMC results for hard spheres \cite{TTS00}. Moreover, an analysis similar to that of Ref.\ \cite{TGS97} can be carried out from the BGK kinetic model. The results though order $\gamma^6$  \cite{TGS99} strongly suggest the asymptotic character of the series \eqref{4.6} and \eqref{4.7}. The theoretical asymptotic analysis of Ref.\ \cite{TGS99} agrees well with a finite-difference numerical solution of the BGK equation \cite{DST99}.

\section{Planar Couette flow}
\label{sec4}
Apart from the Fourier flow, the steady planar Couette flow is perhaps the most basic nonequilibrium state. It corresponds to a fluid enclosed between two infinite parallel plates maintained in relative motion [see Fig.\ \ref{sketch1}(b)]. The plates can be kept at the same temperature or, more generally,  at two different temperatures. In either case, in addition to a velocity profile $\mathbf{u}=u_x(z)\widehat{\mathbf{x}}$, a temperature profile $T(z)$ appears in the system to produce the non-uniform heat flux compensating for the viscous heating. As a consequence, there are two main hydrodynamic lengths: the one associated with the thermal gradient (as in the Fourier flow), i.e., $\ell_{T}=(\partial_z\ln T)^{-1}$, plus the one associated with the shear rate, namely $\ell_{u}=\sqrt{2k_BT/m}(\partial u_x/\partial z)^{-1}$. Therefore, two independent (local) Knudsen numbers can be defined: the reduced thermal gradient defined by Eq.\ \eqref{3.2} and the reduced shear rate
\beq
a=\frac{1}{\lambda_{02}}\frac{\partial u_x}{\partial z},
\label{4.13}
\eeq
where for convenience here we use  the collision frequency $\lambda_{02}$ associated with the viscosity, while in Eq.\ \eqref{3.2} use is made of the collision frequency $\lambda_{11}$ associated with the thermal conductivity.
Like in the planar Fourier flow, the Boltzmann equation of the problem reduces to Eq.\ \eqref{3.0}.  The conservation of momentum implies Eq.\ \eqref{4.4a} as well as
\beq
P_{xz}=\text{const}.
\label{4.14}
\eeq
However, in contrast to Eq.\   \eqref{4.5}, now the  energy conservation equation becomes
\beq
\frac{\partial q_z}{\partial z}+P_{xz}\frac{\partial u_x}{\partial z}=0.
\label{4.15}
\eeq

In 1981, Makashev and Nosik \cite{MN81,N83} extended to the planar Couette flow the solution found in Ref.\ \cite{AMN79} for the planar Fourier flow. More specifically, they proved that a \emph{normal} solution to Eq.\ \eqref{3.0} for Maxwell molecules exists characterized by a constant pressure, i.e., Eq.\ \eqref{3.1a}, and velocity and temperature profiles satisfying
\begin{subequations}
\label{4.16}
\beq
T(z)\frac{\partial{u}_x(z)}{\partial z}=\text{const},
\label{4.16a}
\eeq
\beq
T(z)
\frac{\partial}{\partial z} T(z)\frac{\partial T(z)}{\partial z}=\text{const}.
\label{4.16b}
\eeq
\end{subequations}
These two equations are the counterparts to Eqs.\ \eqref{u=0} and \eqref{3.1b}, respectively. Since $\lambda_{02}\propto p/T$, Eq.\ \eqref{4.16a}  implies that the dimensionless local shear rate defined by Eq.\ \eqref{4.13} is actually uniform across the system, i.e.,
\beq
a=\text{const}.
\label{4.17}
\eeq
In the NS description, Eq.\ \eqref{4.17} follows immediately from the constitutive equation \eqref{2.19} and the exact conservation laws  \eqref{4.4a} and \eqref{4.14}. Here, however, Eq.\ \eqref{4.17} holds even when both Knudsen numbers $\epsilon$ and $a$ are not small and Newton's law \eqref{2.19} fails. This failure can be characterized by means of a nonlinear (dimensionless) shear viscosity $\eta^*(a)$ and two (dimensionless) normal stress differences $\Delta_{1,2}(a)$ defined by
\beq
P_{xz}=-\eta^*(a)\eta_\text{NS}\frac{\partial u_x}{\partial z},
\label{4.18}
\eeq
\beq
\frac{P_{xx}-P_{zz}}{p}={\Delta_1(a)},\quad
\frac{P_{yy}-P_{zz}}{p}={\Delta_2(a)}.
\label{4.19}
\eeq
Note that Eqs.\ \eqref{4.4a}, \eqref{3.1a}, \eqref{4.14}, and \eqref{4.17} are consistent with Eqs.\ \eqref{4.18} and \eqref{4.19}, even though $\eta^*(a)\neq 1$ and $\Delta_{1,2}(a)\neq 0$.

As for Eq.\ \eqref{4.16b}, it can also be justified at the NS level by the constitutive equation \eqref{2.20} and the exact conservation equation \eqref{4.15}. Again, the validity of Eq.\ \eqref{4.16b} goes beyond the scope of Fourier's law \eqref{2.20}. More specifically,  Eq.\ \eqref{4.15} yields
\beq
\frac{1}{\lambda_{11}}\frac{\partial q_z}{\partial z}=\frac{3p}{2}\eta^*(a)a^2=\text{const}.
\label{4.20}
\eeq
Equations \eqref{4.16b} and \eqref{4.20} are consistent with a heat flux component  $q_z$  given by a modified Fourier's law of the form
\beq
q_z=-\kappa^*(a)\kappa_\text{NS}\frac{\partial T}{\partial z},
\label{4.21}
\eeq
where $\kappa^*(a)\neq 1$ is a nonlinear (dimensionless) thermal conductivity.
{}Equations \eqref{4.20} and \eqref{4.21} allow us to identify the constant on the right-hand side of Eq.\ \eqref{4.16b}, namely
\beq
T
\frac{\partial}{\partial z} T\frac{\partial T}{\partial z}=-\frac{3m (\lambda_{11}T)^2}{5k_B}a^2\theta(a)=\text{const},
\label{4.22}
\eeq
where $\theta(a)\equiv \eta^*(a)/\kappa^*(a)$ is a sort of nonlinear Prandtl number.
Making use of Eq.\ \eqref{3.2}, Eq.\ \eqref{4.22} can be rewritten as
\beq
\sqrt{T}\epsilon\frac{\partial}{\partial T}\left(\sqrt{T}\epsilon\right)=-\frac{6}{5}a^2\theta(a).
\label{4.26}
\eeq

To prove the consistency of the assumed hydrodynamic profiles \eqref{3.1a} and \eqref{4.16} we can proceed along similar lines as in the case of the planar Fourier flow \cite{GS03,TS95}. First, we introduce the dimensionless velocity distribution function
\beq
\phi(\mathbf{c};\epsilon,a)=\frac{1}{n(z)}\left[\frac{2k_BT(z)}{m}\right]^{3/2}f(z,\mathbf{v}),
\label{4.23}
\eeq
where now it is important to notice that the definition  \eqref{1.4:23} includes the flow velocity $\mathbf{u}$. Therefore,
\beq
\frac{\partial f}{\partial z}=\frac{\partial T}{\partial z}\frac{\partial f}{\partial T}+ \frac{\partial u_x}{\partial z}\frac{\partial f}{\partial u_x},
\label{4.24}
\eeq
where
\beq
\frac{\partial f}{\partial u_x}=-n\left(\frac{m}{2k_BT}\right)^2\frac{\partial \phi}{\partial c_x}.
\label{4.25}
\eeq
The derivative $\partial f/\partial T$ is given again by Eq.\ \eqref{3.3.n}, but now $\partial \epsilon/\partial T=-T^{-1}\left(\frac{1}{2}\epsilon+\frac{6}{5}\epsilon^{-1}a^2\theta\right)$ on account of Eq.\ \eqref{4.26}. In summary,
\beq
\frac{\partial}{\partial z}
f(z,\mathbf{v})=-n\left(\frac{m}{2k_BT}\right)^2\lambda_{11}\left[\frac{\epsilon}{2}\left(2+
\frac{\partial}{\partial\mathbf{c}}\cdot\mathbf{c}+\epsilon\frac{\partial}{\partial
\epsilon}\right)+\frac{6}{5}a^2\theta(a)\frac{\partial}{\partial
\epsilon}+\frac{3}{2}a\frac{\partial}{\partial c_x}\right]\phi(\mathbf{c};\epsilon,a).
\label{4.27}
\eeq
Consequently,  the Boltzmann equation (\ref{3.0}) becomes
\beqa
-c_z\left[\frac{\epsilon}{2}\left(2+
\frac{\partial}{\partial\mathbf{c}}\cdot\mathbf{c}+\epsilon\frac{\partial}{\partial
\epsilon}\right)\right.&+&\left.\frac{6}{5}a^2\theta(a)\frac{\partial}{\partial
\epsilon}+\frac{3}{2}a\frac{\partial}{\partial c_x}\right]\phi(\mathbf{c};\epsilon,a)\nn&=& \frac{n}{\lambda_{11}} Q\int
d\mathbf{c}_1\int
d\Omega\,\mathcal{B}(\chi)\left[\phi(\mathbf{c}_1';\epsilon,a)\phi(\mathbf{c}';\epsilon,a)-\phi(\mathbf{c}_1;\epsilon,a)\phi(\mathbf{c};\epsilon,a)\right]\nn
&\equiv& \overline{J}[\mathbf{c}|\phi(\epsilon,a),\phi(\epsilon,a)].
\label{4.28}
\eeqa
Of course,  Eq.\ \eqref{4.28} reduces to Eq.\ \eqref{2.5.n} in the special case of the Fourier flow ($a=0$).

For simplicity, we consider now the following non-orthogonal moments of order $\kk=2\rr+\ell$,
\beq
\overline{M}_{\rr\ell{h}}(\epsilon,{a})=\int d\mathbf{c}\,c^{2\rr}c_z^{\ell{-h}}{c_x^h}
\phi(\mathbf{c};\epsilon,{a}),\quad 0\leq h\leq\ell,
\label{4.29}
\eeq
instead of the orthogonal moments \eqref{2.15}. By definition, $\overline{M}_{000}=1$, $\overline{M}_{010}=\overline{M}_{011}=0$, and $\overline{M}_{100}=\frac{3}{2}$. According to Eq.\ \eqref{4.28} the moment equations read
\beqa
\left[\frac{\epsilon}{2}\left(2\rr+\ell-1-\epsilon\frac{\partial}{\partial
\epsilon}\right)\right.&-&\left.\frac{6}{5}a^2\theta(a)\frac{\partial}{\partial
\epsilon}\right]\overline{M}_{\rr,\ell+1,{h}}+\frac{3}{2}a\left(2\rr \overline{M}_{\rr-1,\ell+2,h+1}+h\overline{M}_{\rr,\ell,h-1}\right)\nn
&=&\int
d\mathbf{c}\,c^{2\rr}c_z^{\ell{-h}}{c_x^h}
 \overline{J}[\mathbf{c}|\phi,\phi]\equiv \overline{J}_{\rr\ell {h}}(\epsilon,{a}).
 \label{4.30}
\eeqa
While this hierarchy is much more involved than Eq.\ \eqref{3.7bis}, it can be easily checked to be consistent with solutions of the form
\beq
\overline{M}_{\rr\ell h}(\epsilon,a)=\sum_{j=0}^{2\rr+\ell-2}\overline{m}_j^{(\rr\ell h)}(a)\epsilon^j,\quad \overline{m}_j^{(\rr\ell h)}(a)=0\text{ if }j+\ell=\text{odd}.
\label{4.31}
\eeq
Therefore, the moments $\overline{M}_{\rr\ell{h}}(\epsilon,{a})$ of order $2\rr+\ell\geq 2$ are again \emph{polynomials} in the thermal Knudsen number $\epsilon$ of degree $2\rr+\ell-2$ and parity $\ell$.
In particular, the moments of second order are independent of $\epsilon$, which yields Eqs.\ \eqref{4.18} and \eqref{4.19} with
\beq
\eta^*(a)=-\frac{2\overline{m}_0^{(021)}(a)}{a},
\label{4.32}
\eeq
\beq
\Delta_1(a)=2\left[\overline{m}_0^{(022)}(a)-\overline{m}_0^{(020)}(a)\right],\quad \Delta_2(a)=3-2\left[\overline{m}_0^{(022)}(a)+2\overline{m}_0^{(020)}(a)\right].
\label{4.33}
\eeq
The third-order moments are linear in $\epsilon$, as anticipated by Eq.\ \eqref{4.21} with
\beq
\kappa^*(a)=-\frac{4}{5}\overline{m}_1^{(110)}(a).
\label{4.34}
\eeq
Besides, the shearing induces a component of the heat flux parallel to the flow but normal to the thermal gradient:
\beq
q_x=\Phi(a)\kappa_\text{NS}\frac{\partial T}{\partial z},\quad \Phi(a)=\frac{4}{5}\overline{m}_1^{(111)}(a).
\label{4.35}
\eeq

Since the reduced shear rate $a$ is constant in the bulk domain, it is not difficult to get the spatial dependence of the dimensional moments
\beqa
M_{\rr\ell h}(z)&=&\int d\mathbf{v}\, |\mathbf{v}-\mathbf{u}(z)|^{2\rr}v_z^{\ell-h} [v_x-u_x(z)]^h f(z,\mathbf{v})\nn
&=& n(z) \left[\frac{2k_BT(z)}{m}\right]^{\rr+\ell/2}\overline{M}_{\rr\ell h}(\epsilon(z),a).
\label{4.36}
\eeqa
Inserting Eq.\ \eqref{4.31} into Eq.\ \eqref{4.36} one has
\beq
M_{\rr\ell h}(z)=\frac{2p}{m}\left(\frac{2k_B}{m}\right)^{\rr+\ell/2-1}\;\sum_{j=0}^{\rr-1+[\ell/2]}\overline{m}_{2\rr+\ell-2-2j}^{(\rr\ell h)}(a)\left[\sqrt{T(z)}\epsilon(z)\right]^{2\rr+\ell-2-2j}\left[T(z)\right]^j.
\label{4.36b}
\eeq
As in the case of the Fourier flow, it is convenient to introduce the scaled spatial variable $s$ through Eq.\ \eqref{3.18}. In terms of this variable, Eqs.\ \eqref{4.16} can be integrated to give
\beq
u_x(s)=\frac{3}{2} a s,\quad T(s)=T_0\left[1+\frac{\epsilon_0}{\sqrt{2k_BT_0/m}}s-\frac{3 m}{10k_BT_0} a^2\theta(a)s^2\right],
\label{4.37}
\eeq
where use has been made of Eqs.\ \eqref{4.13} and \eqref{4.22}. Thus,  $u_x$ and $T$ are linear and quadratic functions of $s$, respectively. Analogously, $\sqrt{T}\epsilon$ is linear in $s$, namely
\beq
\sqrt{T(s)}\epsilon(s)=\sqrt{T_0}\epsilon_0-\frac{3\sqrt{2m/k_B}}{5}a^2\theta(a) s.
\label{4.38}
\eeq
Therefore, in view of Eq.\ \eqref{4.36b}, it turns out that, if expressed in terms of $s$, the dimensional moment $M_{\rr\ell h}(s)$ is a polynomial of degree $2\rr+\ell -2$. Inverting Eq.\ \eqref{3.18} one gets the relationship between $s$ and $z$:
\beq
z=\ell_\text{mfp}(0)\frac{s}{\sqrt{2k_BT_0/m}}\left[1+\frac{\epsilon_0}{2}\frac{s}{\sqrt{2k_BT_0/m}}-a^2\theta(a)\frac{s^2}{10 k_BT_0/m}\right].
\label{4.39}
\eeq
The solution to this cubic equation gives $s$ as a function of $z$.
Of course, Eqs.\ \eqref{3.19} and \eqref{3.20} are recovered from Eqs.\ \eqref{4.39} and \eqref{4.37}, respectively, by setting $a=0$.

The key difference with respect to the Fourier flow case [cf.\ Eq.\ \eqref{3.14}] is that, while the coefficients $\overline{m}_{j}^{(\rr\ell)}$ were pure numbers, now the coefficients $\overline{m}_j^{(\rr\ell h)}(a)$, as well as $\theta(a)$, are nonlinear functions of the shear Knudsen number $a$ (of parity equal to $h$). Unfortunately, the full dependence $\overline{m}_j^{(\rr\ell h)}(a)$ cannot be recursively obtained from Eq.\ \eqref{4.30} because the term headed by $a^2\theta(a)$ couples $\overline{m}_j^{(\rr\ell h)}(a)$ to $\overline{m}_{j+1}^{(\rr,\ell+1, h)}(a)$. On the other hand, since that coupling is at least of order $a^2$, it is possible to get recursively the numerical coefficients of the expansions of $\overline{m}_j^{(\rr\ell h)}(a)$ in powers of $a$ \cite{TS95}:
\beq
\overline{m}_j^{(\rr\ell h)}(a)=\sum_{i=0}^\infty \overline{m}_{ji}^{(\rr\ell h)}a^i,\quad \overline{m}_{ji}^{(\rr\ell h)}=0\text{ if }i+h=\text{odd}.
\label{4.40}
\eeq
Obviously, the coefficients $\overline{m}_{j0}^{(\rr\ell h)}$ are those of the Fourier flow and are obtained from the scheme of Fig.\ \ref{arrows}. Their knowledge allows one to get $\overline{m}_{j1}^{(\rr\ell h)}$, and so on. In general, the coefficients $\overline{m}_{ji}^{(\rr\ell h)}$ can be determined from the previous knowledge of the coefficients $\overline{m}_{j'i'}^{(\rr'\ell' h')}$ with $2\rr'+\ell'\leq 2\rr+\ell$ and $2\rr'+\ell'+j'+i'\leq 2\rr+\ell+j+i$. {}From the results to order $a^3$ one gets \cite{GS03,TS95}
\beq
\eta^*(a)
=1+\eta_2^*
a^2+
\mathcal{O}
(a^4),\quad \eta_2^*=-\frac{149}{45},
\label{4.41}
\eeq
\beq
\kappa^*(a)=1+\kappa_2^* a^2+\mathcal{O}(a^4),\quad \kappa_2^*=-\frac{15238 + 45 A_4/A_2}{2100},
\label{4.42}
\eeq
\beq
\theta(a)=1+ \theta_2 a^2+\mathcal{O}(a^4),\quad \theta_2=\frac{24854 + 135A_4/A_2}{6300},
\label{4.43}
\eeq
\beq
\Delta_1(a)
=\frac{14}{5}
a^2+
\mathcal{O}
(a^4),\quad \Delta_2(a)
=\frac{4}{5}
a^2+
\mathcal{O}
(a^4),
\label{4.44}
\eeq
\beq
\Phi(a)
=\frac{7}{2}
a+
\mathcal{O}
(a^3).
\label{4.45}
\eeq
The coefficients in Eqs.\ \eqref{4.44} and \eqref{4.45} are of Burnett order, while those in Eqs.\ \eqref{4.41}--\eqref{4.43} are of super-Burnett order.
Table \ref{tab:3} compares the latter coefficients with the predictions of the BGK kinetic model \cite{BSD87,D90,GS03,KDSB89b}, the Ellipsoidal Statistical (ES) kinetic model \cite{GH97,MG98} first proposed by Holway \cite{ATPP00,C88,H66}, Grad's 13-moment method \cite{RC97,RC98}, and the regularized 13-moment (R13) method \cite{S05,ST08,TTS09}. Despite their simplicity, the BGK and ES kinetic models provide reasonable values. The 13-moment method, however, gives wrong signs for $\kappa_2^*$ and $\theta_2$. On the other hand, the more elaborate R13-moment method practically predicts the right results.

\begin{table}[t]
\caption{Planar Couette flow. Numerical values of the super-Burnett coefficients $\eta_2^*$, $\kappa_2^*$, and $\theta_2=\eta_2^*-\kappa_2^*$, according to the Boltzmann equation (for the IPL Maxwell potential) and several approximations}
\centering
\label{tab:3}       
\begin{tabular}{cccccc}
\hline\noalign{\smallskip}
Coefficient & Boltzmann&BGK model&Ellipsoidal Statistical model&13-moment&R13-moment\\[3pt]
\tableheadseprule\noalign{\smallskip}
$\eta_2^*$&$-3.311$&$-3.60$&$-4.20$&$-2.60$&$-3.311$\\
$\kappa_2^*$&$-7.260$&$-6.48$&$-7.88$&$0.42$&$-7.256$\\
$\theta_2$&$3.948$&$2.88$&$3.68$&$-3.02$&$3.945$\\
\noalign{\smallskip}\hline
\end{tabular}
\end{table}

\begin{figure*}
\centering
  \includegraphics[width=0.7\textwidth]{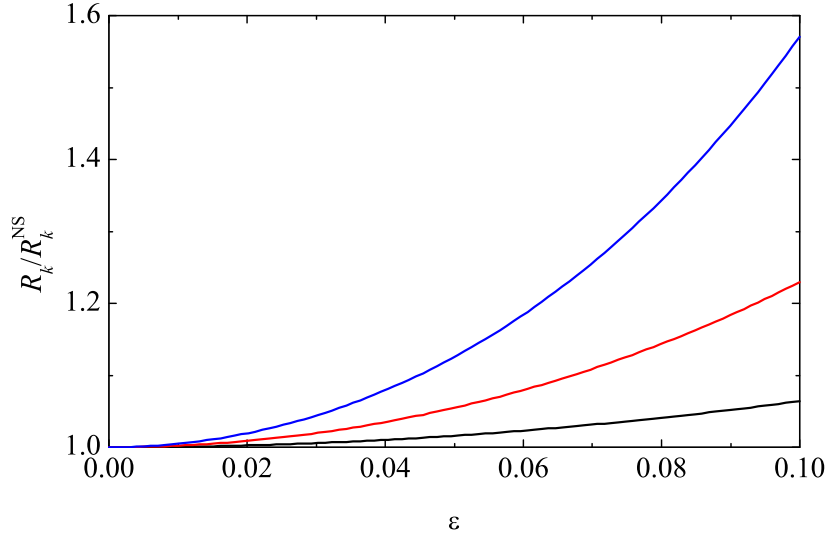}
\caption{Planar Couette flow. Plot of the ratio $R_\rr(\epsilon)/R_\rr^\text{NS}$ for, from bottom to top, $\rr=1$, 2,  and 3.}
\label{Rr}       
\end{figure*}

The quantities \eqref{4.18}, \eqref{4.19}, \eqref{4.21}, and \eqref{4.35} correspond to moments of second and third order. As an illustration of higher-order moments, let us introduce the quantities
\beqa
R_{\rr}(\epsilon)&\equiv& \lim_{a\to 0}\frac{\overline{M}_{\rr21}(\epsilon,a)}{\overline{M}_{021}(a)}=\lim_{a\to 0}\frac{\langle c^{2\rr} c_x c_z\rangle}{\langle  c_x c_z\rangle}\nn
&=&\frac{1}{\overline{m}_{01}^{(021)}}\sum_{j=0}^{\rr}{\overline{m}_{2j,1}^{(\rr21)}}\epsilon^{2j}.
\label{4.46}
\eeqa
The functions $R_{\rr}(\epsilon)$ for $\rr=1$, 2, and 3 are listed in Table \ref{tab:4}, while  Fig.\ \ref{Rr} shows the ratio $R_\rr(\epsilon)/R_\rr^\text{NS}$, where $R_\rr^\text{NS}=R_\rr(0)$ is the NS value.

\begin{table}[t]
\caption{Planar Couette flow. Scaled moments $R_\rr(\epsilon)$, as defined by Eq.\ \protect\eqref{4.46}, for $\rr=1$, 2, and 3}
\centering
\label{tab:4}       
\begin{tabular}{cc}
\hline\noalign{\smallskip}
$\rr$ &$R_\rr(\epsilon)$\\[3pt]
\tableheadseprule\noalign{\smallskip}
$1$&$\frac{7}{2} + 22.41 \epsilon^2$\\
$2$&$\frac{63}{4} + 339.8 \epsilon^2 + 2.184\times 10^3 \epsilon^4$\\
$3$&$\frac{693}{8} + 4.167\times 10^3 \epsilon^2 + 7.153\times 10^4 \epsilon^4 +
 5.977 \times 10^5 \epsilon^6$\\
\noalign{\smallskip}\hline
\end{tabular}
\end{table}

Analogously to the case of the Fourier flow, a comparison between the theoretical expressions of the orthogonal moments related to $R_\rr(\epsilon)$ for $\rr=1$--$3$ and the numerical values obtained from the DSMC method for $\epsilon\lesssim 0.08$ presents an excellent agreement \cite{GTRTS06,GTRTS07}.
Although the full dependence of the moments on the reduced shear rate $a$ cannot be evaluated exactly from the Boltzmann equation, this goal has been achieved in the case of the BGK \cite{BSD87,GS03,KDSB89b} and ES \cite{GH97,GS03} kinetic models, where one can also obtain the velocity distribution function $\phi(\mathbf{c};\epsilon,a)$ itself. The results show that the CE expansion of the distribution function in powers of both $\epsilon$ and $a$ is only asymptotic. The BGK and ES predictions for the coefficients $\eta^*(a)$, $\kappa^*(a)$, $\theta(a)$, $\Delta_{1,2}(a)$, and $\Phi(a)$ compare favorably well with DSMC results for both Maxwell molecules and hard spheres \cite{MSG00}. The BGK kinetic model has also been used to assess the influence of an external body force $\mathbf{F}=-mg\widehat{\mathbf{z}}$ on the transport properties of the Couette flow \cite{TGS99b}. The results show that such an influence tends to decrease as the shear rate increases in the case of the rheological properties $\eta^*(a)$ and $\Delta_{1,2}(a)$, while it
is especially important in the case of the heat flux coefficient $\Phi(a)$.

\section{Force-driven Poiseuille flow}
\label{sec5}
One of the most well-known textbook examples in fluid dynamics is the Poiseuille flow \cite{T88}. It consists of the steady flow along a channel of constant cross section produced by a pressure difference at the distant ends of the channel. At least at the NS order,  essentially the same type of flow can be generated
 by the action of a uniform longitudinal body force $\mathbf{F}=-mg\widehat{\mathbf{z}}$
(e.g., gravity) instead of a longitudinal pressure gradient. This force-driven Poiseuille flow [see Fig.\ \ref{sketch1}(c)]
has received much attention from  computational
\cite{KMZ87,KMZ89,MBG97,RC98b,TE97,TTE97,ZGA02} and  theoretical
\cite{AS92,ATN02,ELM94,GVU08,HM99,RC98b,STS03,ST06,ST08,TTS09,TSS98,TS94,TS01,TS04,UG99b,X03} points
of view. This interest has been mainly fueled by the fact that the
force-driven Poiseuille flow provides a nice example illustrating
the limitations of   the NS description
 in the bulk domain (i.e., far away from the boundary layers).

The Boltzmann equation for this problem  becomes
\begin{equation}
v_x \frac{\partial }{\partial x} f(x,\mathbf{v}) - g \frac{\partial }{\partial
v_z} f(x,\mathbf{v})= J \lbrack \mathbf{v}| f, f \rbrack .
\label{5.22}
\end{equation}
The apparent similarity  to Eq.\ \eqref{4.2} is deceptive: although the velocity distribution function only depends on the coordinate ($x$) orthogonal to the plates, this axis is perpendicular to the force, in contrast to the case of Eq.\ \eqref{4.2}. The conservation laws of momentum and energy imply
\beq
{ P_{xx}}=\text{const},\quad  \frac{\partial
P_{xz}}{\partial x}=-\rho  g, \quad \frac{\partial q_{x}}{\partial x}+P_{xz}\frac{\partial
u_z}{\partial x}=0.
\label{5.3}
\eeq
The appropriate (dimensional) moments are defined similarly to the first line of Eq.\ \eqref{4.36}, namely
\beq
M_{\rr\ell h}(x)=\int d\mathbf{v}\, |\mathbf{v}-\mathbf{u}(x)|^{2\rr}v_x^{\ell-h} [v_z-u_z(x)]^h f(x,\mathbf{v}), \quad 0\leq h\leq \ell.
\label{5.1}
\eeq
Because of the symmetry properties of the problem, $u_z(x)$ is an even function of $x$ and $M_{\rr\ell h}(x)$ is an even (odd) function of $x$ if $\ell-h$ is even (odd). Seen as functions of $g$, $u_z(x)-u_z(0)$ is an odd function, while $M_{\rr\ell h}(x)$ is an even (odd) function if $h$ is even (odd). The corresponding hierarchy of moment equations reads
\beqa
\frac{\partial}{\partial x} M_{\rr,\ell+1,h}(x)&+&\frac{\partial u_z(x)}{\partial x}\left[2\rr M_{\rr-1,\ell+2,h+1}(x)+hM_{\rr,\ell,h-1}(x)\right]\nn
&+&g\left[2\rr M_{\rr-1,\ell+1,h+1}(x)+hM_{\rr,\ell-1,h-1}(x)\right]=J_{\rr\ell h}(x).
\label{5.2}
\eeqa
Again, moments of order $2\rr+\ell$ are coupled to moments of order $2\rr+\ell+1$, and so their full dependence  on $x$ and $g$ cannot be determined, even in the bulk domain and for Maxwell models. However, the problem
can be solved by a recursive scheme~\cite{TSS98} if the moments
are expanded in powers of $g$ around a reference equilibrium state
parameterized by $u_z(0)=u_0$, $p(0)=p_0$, and $T(0)=T_0$:
\beq
u_z(x)=u_0+ u_z^{(1)}(x) g+ u_z^{(3)}(x) g^3+\cdots,\quad
p(x)=p_0+ p^{(2)}(x) g^2+ p^{(4)}(x) g^4+\cdots,
\label{5.n4.1}
\eeq
\beq
T(x)=T_0+ T^{(2)}(x) g^2+ T^{(4)}(x) g^4+\cdots,\quad
M_{\rr \ell h}(x)=M_{\rr \ell h}^{(0)}+M_{\rr \ell h}^{(1)}g+M_{\rr \ell h}^{(2)}g^2+\cdots.
\label{5.n4.4}
\eeq
 Due
to symmetry reasons, $u_z^{(j)}(x)=0$ if $j=\text{even}$,
$p^{(j)}(x)=T^{(j)}(x)=0$ if $j=\text{odd}$, and
$M_{\rr \ell h}^{(j)}(x)=0$ if $h+j=\text{odd}$. The functions $u_z^{(j)}(x)$, $p^{(j)}(x)$, and $T^{(j)}(x)$ are even, while $M_{\rr \ell h}^{(j)}(x)$ has the same parity as $\ell-h$. In fact, it turns out that those functions are just \emph{polynomials} of degree $2j$ (or $2j-1$ if $\ell-h=\text{odd}$) \cite{TSS98}, namely
\beq
u_z^{(j)}(x)=\sum_{i=1}^j u_z^{(j,2i)}x^{2i},\quad
p^{(j)}(x)=\sum_{i=2}^j p^{(j,2i)}x^{2i},
\label{5.n4.5}
\eeq
\beq
T^{(j)}(x)=\sum_{i=2}^j T^{(j,2i)}x^{2i}\;,\quad
M_{\rr\ell h}^{(j)}(x)=\sum_{i=0}^{2j}m_{\rr\ell h}^{(ji)}x^i,
\label{5.n4.6}
\eeq
where $m_{\rr\ell h}^{(ji)}=0$ if $i+\ell-h=\text{odd}$ and $m_{\rr\ell h}^{(12)}=0$. The numerical
coefficients $u_z^{(j,2i)}$, $p^{(j,2i)}$, $T^{(j,2i)}$,
and $m_{\rr\ell h}^{(ji)}$ are  determined recursively by
inserting (\ref{5.n4.1})--(\ref{5.n4.6}) into (\ref{5.2}) and equating
the coefficients of the same powers in $g$ and $x$  in both sides.
This yields a hierarchy of \textit{linear} equations for the
unknowns. This rather cumbersome scheme has been solved through
order $g^2$ in Ref.\ \cite{TSS98}. The results for the hydrodynamic
profiles and the fluxes are
\beq
u_z(x)=u_0+\frac{\rho_0 g}{2\eta_0}{x}^2+\mathcal{O}(g^3),\quad
p(x)=p_0\left[1+C_p\left(\frac{mg}{k_BT_0}\right)^2x^2\right]+ \mathcal{
O}(g^4),
\label{5.3.9}
\eeq
\beq
T(x)=T_0\left[1-\frac{\rho_0^2 g^2}{12\eta_0\kappa_0T_0}{x}^4+
C_T\left(\frac{mg}{k_B T_0}\right)^2x^2\right] +\mathcal{ O}(g^4),
\label{5.3.11}
\eeq
\beq
P_{xx}=p_0\left(1-C_{xx}\frac{\rho_0\eta_0^2g^2}{p_0^3}\right)+
{\cal O}(g^4),
\label{5.3.14}
\eeq
\beq
P_{zz}(x)=p_0\left[1+\frac{7}{3}C_p
\left(\frac{mg}{k_BT_0}\right)^2x^2+C_{zz}\frac{\rho_0\eta_0^2g^2}{p_0^3}\right]+
\mathcal{O}(g^4),
\label{5.3.13}
\eeq
\beq
P_{xz}(x)=-\rho_0gx\left[1+\frac{\rho_0^2
g^2}{60\eta_0\kappa_0T_0}{x}^4+\frac{C_p-C_T}{3}
\left(\frac{mg}{k_BT_0}\right)^2x^2\right]+ \mathcal{ O}(g^5),
\label{5.3.16}
\eeq
\beq
q_x(x)=\frac{\rho_0^2g^2}{3\eta_0}x^3+{\cal O}(g^4),\quad
q_z(x)=C_q{m g \kappa_0} +{\cal O}(g^3),
\label{5.3.18}
\eeq
where $\rho_0=\rho(0)=mp_0/k_BT_0$, $\eta_0=\eta_\text{NS}(0)$, and $\kappa_0=\kappa_\text{NS}(0)$.
The numerical values of the coefficients $C_p$, $C_T$, $C_{xx}$,
$C_{zz}$, and $C_q$ are shown in Table \ref{tab:5}, which also includes the values predicted by the NS constitutive equations, the Burnett equations \cite{UG99b}, Grad's 13-moment method \cite{RC98b}, the R13-moment method \cite{ST08,TTS09}, a 19-moment method \cite{HM99}, and the BGK kinetic model \cite{TS94}.

\begin{table}[t]
\caption{Force-driven Poiseuille flow. Numerical values of the coefficients $C_p$, $C_T$, $C_{xx}$,
$C_{zz}$, and $C_q$, according to the Boltzmann equation (for the IPL Maxwell potential) and several approximations}
\centering
\label{tab:5}       
\begin{tabular}{cccccccc}
\hline\noalign{\smallskip}
Coefficient & Boltzmann&NS&Burnett&13-moment&R13-moment&19-moment&BGK model\\[3pt]
\tableheadseprule\noalign{\smallskip}
$C_p$&$1.2$& $0$ &$1.2$&$1.2$&$1.2$&$1.2$&$1.2$ \\
$C_T$& $1.0153$& $0$ &$0$&$0.56$&$0.9295$&$1.04$&$0.76$\\
$C_{xx}$ &$6.2602$& $0$ & $0$&$0$&$3.36$&--&$12.24$\\
$C_{zz}$ &$6.4777$& $0$ &$0$& $0$&$3.413$&--&$13.12$\\
$C_q$ & $0.4$& $0$ &$0.4$&$0.4$&$0.4$&$0.4$&$0.4$\\
\noalign{\smallskip}\hline
\end{tabular}
\end{table}

\begin{figure*}
\centering
  \includegraphics[width=\textwidth]{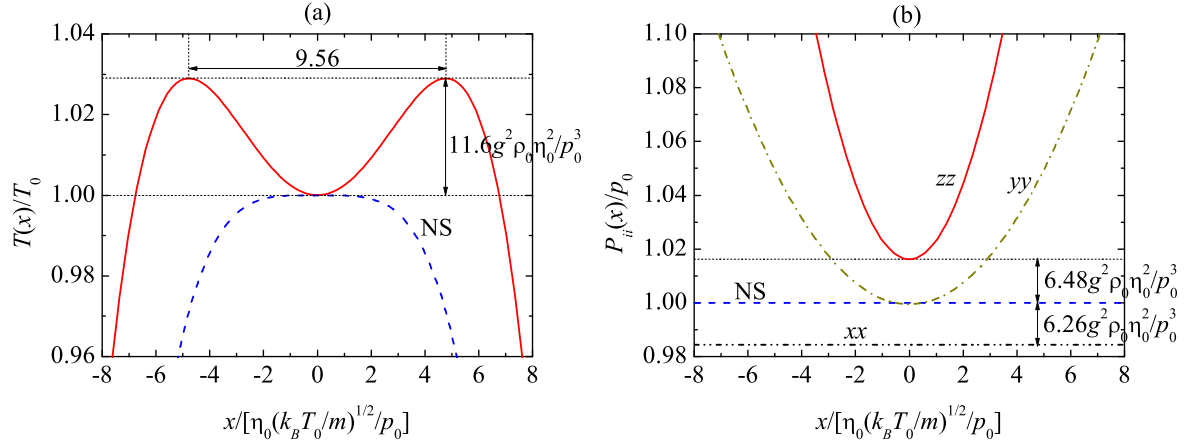}
\caption{Force-driven Poiseuille flow. Plot of ({a}) $T(x)/T_0$ and ({b})
$P_{ii}(x)/p_0$ for $g=0.05  p_0^{3/2}/\rho_0^{1/2}\eta_0$, according to the NS equations
 and the
Boltzmann equation for Maxwell molecules.}
\label{Poise}       
\end{figure*}

All the coefficients $C_p$, $C_T$, $C_{xx}$,
$C_{zz}$, and $C_q$ vanish in the NS description \cite{ST06}. At a qualitative level, the main correction to the NS results appears for
the temperature profile. While, according to the NS equations, the
temperature has a maximum at the mid plane $x=0$, the Boltzmann
equation shows that, because of the extra quadratic term headed by
$C_T$, the temperature actually presents a local minimum $T_0$ at
$x=0$ surrounded by two symmetric maxima $T_{\text{max}}$ at $x=\pm
x_{\text{max}}$, where $x_{\text{max}}\equiv
\sqrt{6C_T\eta_0\kappa_0T_0}/p_0$. The relative height of the maxima
is $(T_\text{max}-T_0)/T_0= 2C_T(mgx_{\text{max}}/2k_BT_0)^2$.
Therefore,  the temperature  does not present a flat maximum at
the middle of the channel but instead exhibits a \textit{bimodal}
shape with a local minimum
 surrounded by two symmetric maxima at a
distance of a few mean free paths. This is illustrated by Fig.\ \ref{Poise}(a) for $g=0.05  p_0^{3/2}/\rho_0^{1/2}\eta_0$.
The Fourier law is dramatically
violated since in the slab enclosed by the two maxima the transverse
component of the heat flux is parallel (rather than anti-parallel)
to the thermal gradient.
Other non-Newtonian effects are the fact that the hydrostatic pressure is not uniform across the system ($C_p\neq 0$) and the existence of normal stress differences ($C_{xx}\neq 0$, $C_{zz}\neq 0$), as illustrated by Fig.\ \ref{Poise}(b) for $g=0.05  p_0^{3/2}/\rho_0^{1/2}\eta_0$. Moreover,
there exists a heat flux component orthogonal to the thermal gradient ($C_q\neq 0$).

The coefficients $C_p=\frac{6}{5}=1.2$ and $C_q=\frac{2}{5}=0.4$ are already captured by the Burnett
description \cite{TSS98,UG99b}. On the other hand, the
determination of $C_T$, $C_{xx}$, and $C_{zz}$ requires the
consideration of, at least, super-Burnett
contributions (e.g., $\partial^3 T/\partial x^3$), as first pointed out in Ref.\ \cite{TSS98}. However, a complete determination of those three coefficients requires to retain super-super-Burnett terms (e.g., $\partial^4 T/\partial x^4$). In general, in order to get the fluxes through order $g^{2j}$, one needs to consider the CE expansion though order $2(j+1)$ in the gradients \cite{TSS98}; but this  would also provide many extra terms of order higher than $g^{2j}$, that should be discarded.  The 13-moment approximation~\cite{RC98b}
is able to predict, apart from the correct Burnett-order coefficients $C_p$ and
$C_q$, a non-zero value $C_T=\frac{14}{25}=0.56$.  The R13-moment approximation significantly improves the value of $C_T$ ($C_T=\frac{1859}{2000}=0.9295$) and accounts for non-zero values of $C_{xx}$ and $C_{zz}$ ($C_{xx}=\frac{84}{25}=3.36$, $C_{zz}=\frac{256}{75}\simeq 3.413$). Although both values are almost half the correct ones,  they satisfactorily show that $C_{zz}$ is only slighter larger than $C_{xx}$, implying that $C_{yy}=-(C_{zz}-C_{xx})$ is rather small.
The more complicated 19-moment approximation~\cite{HM99}
gives  $C_T=\frac{26}{25}=1.04$ for Maxwell
molecules but the predictions for $C_{xx}$ and $C_{zz}$ were not explicitly given in Ref.\ \cite{HM99}.

The solution to the BGK equation for the
plane Poiseuille flow has been explicitly obtained through order
$g^5$~\cite{TS94}. The results strongly suggest that the series
expansion is only asymptotic, so that from a practical point of view
one can focus on the first few terms. The results agree with the
profiles (\ref{5.3.9})--(\ref{5.3.18}), except that the numerical values
of the coefficients $C_T=\frac{19}{25}=0.76$ and, especially,  $C_{xx}=\frac{306}{25}=12.24$ and $C_{zz}=\frac{328}{25}=13.12$ differ from
those derived from the Boltzmann equation for Maxwell molecules, as
shown in Table \ref{tab:5}.  Interestingly enough, the BGK value of
$C_T$ agrees quite well with DSMC
simulations of the Boltzmann equation for hard spheres \cite{MBG97}.
It is worth mentioning that an exact, non-perturbative solution of the BGK kinetic model exists for the particular value $g=2.5240  p_0^{3/2}/\rho_0^{1/2}\eta_0$ \cite{AS92}.

\section{Uniform shear flow}
\label{sec6}
The uniform shear flow is a time-dependent state generated by the application of Lees--Edwards boundary conditions \cite{LE72}, which are a generalization of the conventional periodic boundary conditions
employed in molecular dynamics of systems in equilibrium. Like the Couette flow, the uniform shear flow can be sketched by Fig.\ \ref{sketch1}(b), except that now  density and  temperature are uniform and the velocity profile is strictly linear \cite{GS03,T56,TM80}. The Boltzmann equation in this state becomes
\beq
\frac{\partial}{\partial t}f(z,\mathbf{v})+v_z\frac{\partial}{\partial z} f(z,\mathbf{v})=J[\mathbf{v}|f,f].
\label{6.0}
\eeq
Since there is no thermal gradient, the only hydrodynamic length is $\ell_{u}=\sqrt{2k_BT/m}(\partial u_x/\partial z)^{-1}$, so that the relevant Knudsen number is the reduced shear rate defined by Eq.\ \eqref{4.13}, which is again constant. On the other hand, in the absence of heat transport, the viscous heating term $P_{xz}\partial u_x/\partial z$ cannot be balanced by the divergence of the heat flux  and so the temperature monotonically increases with time. In other words, the energy balance equation \eqref{4.15} is replaced by
\beq
\frac{3nk_B}{2}\frac{\partial T}{\partial t}+P_{xz}\frac{\partial u_x}{\partial z}=0.
\label{6.1}
\eeq
We can still define the dimensionless distribution function \eqref{4.23} and the dimensionless moments \eqref{4.29}, but now $\epsilon=0$ (no thermal gradient), so  these quantities are nonlinear functions of the reduced shear rate $a$ only. The moment hierarchy is formally similar to Eq.\ \eqref{4.30}, except that the first term on the left-hand side is replaced by a term coming from the time dependence of temperature, in the same way as the first term on the left-hand side of Eq.\ \eqref{4.15} is replaced by that of Eq.\ \eqref{6.1}. More explicitly, the moment equations are
\beq
\left(\rr+\frac{\ell}{2}\right)\xi \overline{M}_{\rr\ell h}{+a\left(2\rr \overline{M}_{\rr-1,\ell+2,h+1}+h\overline{M}_{\rr,\ell,h-1}\right)}
=\frac{1}{\lambdabar_{02}} \overline{J}_{\rr\ell {h}}({a}),
 \label{6.2}
\eeq
where $\xi\equiv \lambda_{02}^{-1}\partial \ln T/\partial t$. In contrast to Eq.\ \eqref{4.30}, the hierarchy \eqref{6.2} only couples moments of the same order $2\rr+\ell$ and of lower order, so it can be solved order by order. Note that moments of odd order (like the heat flux) vanish because of symmetry.

\begin{figure*}
\centering
  \includegraphics[width=0.7\textwidth]{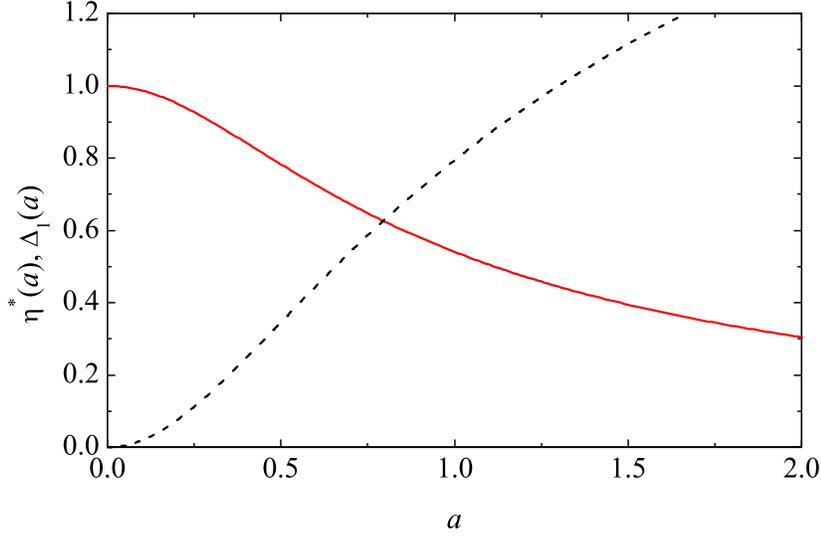}
\caption{Uniform shear flow. Plot of the nonlinear shear viscosity $\eta^*(a)$ (solid line) and  the normal stress difference $\Delta_1(a)$ (dashed line).}
\label{USF}       
\end{figure*}

Specifically, setting $(\rr,\ell,h)=(1,0,0)$, $(0,2,0)$, $(0,2,1)$, and $(0,2,0)$ in Eq.\ \eqref{6.2}, one gets
\beq
\overline{M}_{020}(a)=\frac{1}{2}\frac{1}{1+\xi(a)},\quad \overline{M}_{021}(a)=-\frac{1}{2}\frac{a}{[1+\xi(a)]^2},\quad \overline{M}_{022}(a)=\frac{1}{2}\frac{1+3\xi(a)}{1+\xi(a)},
\label{6.3}
\eeq
where $\xi(a)$ is the real root of the cubic equation $3\xi(1+\xi)^2=2a^2$, namely
\beq
\xi(a)=\frac{4}{3}\sinh^2\left[\frac{1}{6}\cosh^{-1}(1+9a^2)\right].
\label{6.4}
\eeq
Upon deriving Eqs.\ \eqref{6.3} and \eqref{6.4} use has been made of the properties $\overline{M}_{100}=\frac{3}{2}$, $\overline{J}_{100}=0$, $\overline{J}_{020}=-\lambdabar_{02}\left(\overline{M}_{020}-\frac{1}{2}\right)$, $\overline{J}_{021}=-\lambdabar_{02}\overline{M}_{021}$, and $\overline{J}_{022}=-\lambdabar_{02}\left(\overline{M}_{022}-\frac{1}{2}\right)$.

\begin{figure*}
\centering
  \includegraphics[width=0.7\textwidth]{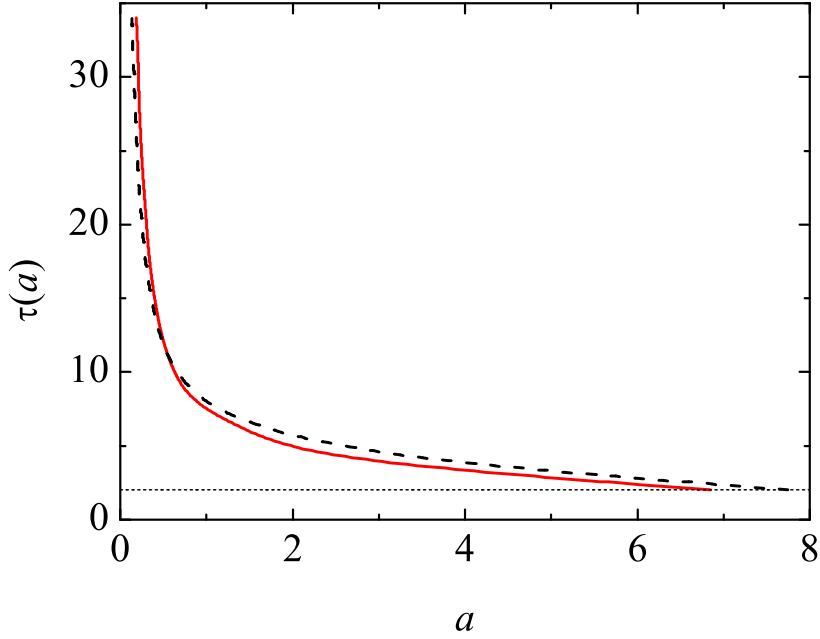}
\caption{Uniform shear flow. Plot of the exponent $\tau(a)$ defined by Eq.\ \protect\eqref{6.6} for the IPL model (solid line) and the VHS model (dashed line).
The dotted horizontal line $\tau=2$ intercepts the curves at $a_c\simeq  6.846$ and $a_c\simeq 7.746$, respectively.}
\label{s(a)}       
\end{figure*}
The nonlinear shear viscosity $\eta^*(a)$, defined by Eq.\ \eqref{4.18},  and the normal stress differences $\Delta_{1,2}(a)$, defined by Eq.\ \eqref{4.19}, are now explicitly given by
\beq
\eta^*(a)=\frac{1}{[1+\xi(a)]^2},\quad \Delta_1(a)=\frac{3\xi(a)}{1+\xi(a)},\quad \Delta_2(a)=0.
\label{6.5}
\eeq
The dependence of $\eta^*(a)$ and $\Delta_1(a)$ on the reduced shear rate $a$ is shown in Fig.\ \ref{USF}. For small shear rates, one has $\xi(a)=\frac{2}{3}a^2+\mathcal{O}(a^4)$, so that $\eta^*(a)=1-\frac{4}{3}a^2+\mathcal{O}(a^4)$, $\Delta_1(a)=2a^2+\mathcal{O}(a^4)$. The latter two quantities differ from the ones in the Couette flow, Eqs.\ \eqref{4.41} and \eqref{4.44}.

Once the second-order moments are fully determined, one can proceed to the fourth-order moments. Although not related to transport properties, they  provide useful information about the population of particles with velocities much larger than the thermal velocity. {}From Eq.\ \eqref{6.2} one gets a closed set of nine linear equations that can be algebraically solved \cite{GS03,SG95,SGBD93}. Interestingly, these moments are well-defined only for shear rates smaller than a certain critical value $a_c$ ($a_c\simeq  6.846$ and $7.746$ for the IPL and VHS models, respectively). Beyond that critical value, the \emph{scaled} fourth-order moments [e.g., $\langle c^4\rangle=\langle V^4\rangle/(2k_BT/m)^2$] monotonically increase in time without upper bound and  diverge in the long-time limit. This clearly indicates that the reduced velocity distribution function exhibits an
algebraic high-velocity tail \cite{GS03,MSG96b,MSG96}
\beq
\phi(\mathbf{c};a)\sim c^{-5-\tau(a)},
\label{6.6}
\eeq
so that those moments of order equal to or larger than $2+\tau(a)$ diverge. In particular, the critical shear rate $a_c$ is the solution to $\tau(a)=2$. The dependence of the exponent $\tau(a)$ on the shear rate is shown in Fig.\ \ref{s(a)}. The scenario described by Eq.\ \eqref{6.6} has been confirmed by DSMC results for Maxwell molecules \cite{MSG97}. On the other hand, hard spheres do not present an algebraic high-velocity tail \cite{MSG96b} and thus all the moments are finite.

\section{Conclusions}
\label{sec7}

In this paper I have presented an overview of a few cases in which the moment equations stemming from the Boltzmann equation for Maxwell molecules can be used to extract exact information about  strongly nonequilibrium properties in the bulk region of the system. This refers to situations where the Knudsen number defined as the ratio between the mean free path and the characteristic distance associated with the hydrodynamic gradients is finite, whilst the  ratio between the mean free path and the size of the system is vanishingly small. In other words, the system size comprises many mean free paths but the hydrodynamic quantities can vary appreciably over a distance on the order of the mean free path.

In the conventional Fourier flow problem, the relevant Knudsen number (in the above sense) is given by Eq.\ \eqref{3.2}. Regardless of the value of $\epsilon$,   it turns out that the infinite moment hierarchy admits a solution where the pressure is uniform, the temperature is linear in the scaled variable $s$ defined by Eq.\ \eqref{3.18} [cf.\ Eq.\ \eqref{3.20}], and the dimensionless moments of order $\kk$ are polynomials in $\epsilon$ of degree $\kk-2$ [cf.\ Eqs.\ \eqref{3.9} and \eqref{3.14}]. The latter implies that the dimensional moments of order $\kk$ are polynomials in $s$ of degree $[\kk/2]-1$ [cf.\ Eq.\ \eqref{3.16}]. In this solution the heat flux is exactly given by Fourier's law, Eq.\ \eqref{NS1}, even at far from equilibrium states.
On the other hand, the situation changes when a gravity field normal to the plates is added. Apart from the Knudsen number $\epsilon$, a relevant parameter is the microscopic Rayleigh number defined by Eq.\ \eqref{4.1}. Assuming $|\gamma|\ll 1$, a perturbation expansion can be carried out about the conventional Fourier flow state, the results exhibiting non-Newtonian effects [cf.\ Eq.\ \eqref{4.10}] and a breakdown of Fourier's law [cf.\ \eqref{4.11}].

When the parallel plates enclosing the gas are in relative motion, one is dealing with the so-called planar Couette flow. The plates can be kept at the same temperature or at different ones. In the former case the state reduces to that of equilibrium when the plates are at rest, while it reduces to the planar Fourier flow in the latter case. In either situation, the shearing produces a non-uniform heat flux to compensate (in the steady state) for the viscous heating, so that a temperature profile is present. This implies that two independent Knudsen numbers can be identified: the one associated with the flow velocity field [cf.\ Eq.\ \eqref{4.13}] and again the one related to the temperature field [cf.\ Eq.\ \eqref{3.2}]. Now the exact solution to the moment equations is characterized by a constant pressure, a constant reduced shear rate $a$, a quadratic dependence of the temperature profile on the scaled spatial variable $s$ [cf.\ Eq.\ \eqref{4.37}], and again a polynomial dependence of the dimensionless moments on the reduced thermal gradient $\epsilon$ [cf.\ Eq.\ \eqref{4.31}]. The dimensional moments of order $\kk$ are polynomials of degree $\kk-2$ in $s$ [cf.\ Eq.\ \eqref{4.36b}]. As a consequence, Newton's law is modified by a nonlinear shear viscosity [cf.\ Eq.\ \eqref{4.18}] and two viscometric functions [cf.\ Eq.\ \eqref{4.19}], and a generalized Fourier's law holds whereby the heat flux is proportional to the thermal gradient but with a nonlinear thermal conductivity [cf.\ Eq.\ \eqref{4.21}] and a new coefficient related to the streamwise component [cf.\ Eq.\ \eqref{4.35}]. The main difficulty lies in the fact that the coefficients of those polynomials are no longer pure numbers but nonlinear function of the reduced shear rate $a$. Therefore, although the structural form of the solution applies for arbitrary $a$, in order to get explicit results one needs to perform a partial CE expansion in powers of $a$ and get the coefficients recursively. In particular, the super-Burnett coefficients $\eta_2^*$ and $\kappa_2^*$ are obtained [cf.\ Eqs.\ \eqref{4.41} and \eqref{4.42}].

One of the nonequilibrium states more extensively studied in the last few years is the force-driven Poiseuille flow. A series expansion about the equilibrium state in powers of the external force allows one to get the hydrodynamic profiles and the moments order by order. Calculations to second order suffice to unveil dramatic deviations from the NS predictions. More explicitly, the temperature profile presents a bimodal shape with a dimple at the center of the channel [cf.\ Fig.\ \ref{Poise}(a)]. Besides, strong normal-stress differences are present [cf.\ Fig.\ \ref{Poise}(b)].

In  the specific states summarized above the moment equations have a hierarchical structure since the divergence of the flux term, represented by $\nabla\cdot\bm{\Phi}_\psi$ in Eq.\ \eqref{2.10}, involves moments of an order higher  than that of $\Psi$. This is what happens in Eqs.\ \eqref{3.7}, \eqref{3.7bis}, \eqref{4.3}, \eqref{4.30}, and \eqref{5.2}. As a consequence, a recursive procedure is needed to obtain explicit results step by step without imposing a truncation closure. On the other hand, the methodology is different (and simpler!) in time-dependent states  that become spatially uniform in the co-moving or Lagrangian frame of reference. In that case the divergence of the flux term $\bm{\Phi}_\psi$    involves moments of the same order as $\psi$ and so the hierarchical character of the moment equations is truncated in a natural way. This class of states include the uniform shear flow \cite{ASB02,GS03,IT56,MS96,MSG96,SG95,SGBD93,T56,TM80} considered in Sec.\ \ref{sec6}, and the uniform longitudinal flow (or homoenergetic extension) \cite{G95,GK96,KDN97,S00a,S03,TM80}. In both cases the moments of second order (pressure tensor) obey a closed set of autonomous equations that can be algebraically solved to get the rheological properties as explicit nonlinear functions of the reduced velocity gradient, the solution exhibiting interesting non-Newtonian effects. Once the moments of second order are determined, one can proceed to the closed set of equations for the moments of fourth order. Here the interesting result is that those moments diverge beyond a certain critical value of the reduced velocity gradient \cite{ASB02,GS03,S00a,SGBD93}, what indicates the existence of an \emph{algebraic} high-velocity tail of the velocity distribution function \cite{ASB02,GS03,MSG96}.

Needless to say, exact results in kinetic theory are important by themselves and also as an assessment of simulation techniques, approximate methods, and kinetic models.
In this sense, it is hoped that the states and results reviewed in this paper can become useful to other researchers.

\begin{acknowledgements}
I am grateful to those colleagues I have had the pleasure to collaborate with in solutions of the moment hierarchy for Maxwell molecules. In particular, I would like to mention (in alphabetical order) J. Javier Brey (Universidad de Sevilla, Spain), James. W. Dufty (University of Florida, USA), Vicente Garz\'o  (Universidad de Extremadura, Spain), and Mohamed Tij (Universit\'e Moulay Isma\"il, Morocco).
This work  has been supported by the
Ministerio de Educaci\'on y Ciencia (Spain) through Grant No.\
FIS2007-60977 (partially financed by FEDER funds) and by the Junta
de Extremadura through Grant No.\ GRU09038.

\end{acknowledgements}

\bibliographystyle{spmpsci}
\bibliography{book_new}   

\end{document}